\documentclass[pdflatex,sn-mathphys-ay]{sn-jnl}

 \usepackage[utf8]{inputenc}
\usepackage[T1]{fontenc}
\usepackage[english]{babel}
\usepackage{comment}
\usepackage{graphicx}%
\usepackage{tabularx}
\usepackage{multirow}%
\usepackage{amsmath,amssymb,amsfonts}%
\usepackage{amsthm}%
\usepackage{mathrsfs}%
\usepackage[title]{appendix}%
\usepackage{xcolor}%
\usepackage{textcomp}%
\usepackage{manyfoot}%
\usepackage{booktabs}%
\usepackage{hyperref,float,array}
\usepackage{bbold} 
 \usepackage{hyperref,float,multirow}
\usepackage{colortbl}
\usepackage{makecell}
\usepackage{booktabs}
\usepackage{listings}%
\usepackage{algorithm,algorithmic}

\newcommand{\BEGIN}{ \STATE \fbox{\textbf{procedure}}}

\newcommand{\Input}{\STATE \textbf{Input}:}
\newcommand{\Output}{\STATE \textbf{Output: }}

\theoremstyle{thmstyleone}%
\theoremstyle{thmstyletwo}%

\theoremstyle{thmstylethree}%

\begin{document}

\title[Article Title]{Generalized random forest for extreme quantile regression}


\author*[1]{\fnm{Lucien M. } \sur{Vidagbandji}}\email{mahutin-lucien.vidagbandji@univ-lehavre.fr}
\author[1]{\fnm{Alexandre} \sur{Berred}}\email{alexandre.berred@univ-lehavre.fr}

\author[2]{\fnm{Cyrille} \sur{Bertelle}}\email{cyrille.bertelle@univ-lehavre.fr}
\author[2]{\fnm{Laurent} \sur{Amanton}}\email{laurent.amanton@univ-lehavre.fr}

	\affil[1]{\orgdiv{Université Le Havre Normandie, Normandie Univ.}, \orgname{LMAH UR 3821}, \orgaddress{ \city{76600},  \state{Le Havre}, \country{France}}}

\affil[2]{\orgdiv{Université Le Havre Normandie, Normandie Univ.}, \orgname{LITIS},\orgaddress{ \city{76600},  \state{Le Havre}, \country{France}}}

\abstract{Quantile regression is a statistical method which, unlike classical regression, aims to predict the conditional quantiles. Classical quantile regression methods face difficulties, particularly when the quantile under consideration is extreme, due to the limited number of data available in the tail of the distribution, or when the quantile function is complex. We propose an extreme quantile regression method based on extreme value theory and statistical learning to overcome these difficulties. Following the Block Maxima approach of extreme value theory, we approximate the conditional distribution of block maxima by the generalized extreme value distribution, with covariate-dependent parameters. These parameters are estimated using a method based on generalized random forests. Applications on simulated data show that our proposed method effectively addresses the mentioned quantile regression issues and highlights its performance compared to other quantile regression approaches based on statistical learning methods. We apply our methodology to daily meteorological data from the Fort Collins station in Colorado (USA).}




\maketitle
\section{Introduction}\label{sec1}
	
Extreme events, defined by their scarcity and potential for severe impact, frequently dominate the news and capture our imagination due to their unpredictability and the losses they cause. Examples of such extreme events include floods, earthquakes, rogue waves, financial crises, to name just a few. Understanding or anticipating the occurrence of these events allows for a deeper analysis of the associated risks. Extreme quantile regression is an appropriate tool in this regard. It is used to model the relationship between explanatory variables (covariates) and the extreme conditional quantiles of a response variable. This approach helps in understanding how factors influence the extreme values of a distribution. Extreme events occur in the tails of distributions, so for risk analysis, it's mainly a question of estimating extreme quantiles, i.e. quantiles at high  level $\tau.$ Specifically, in the univariate context, if $Y\in \mathcal{Y}\subset\mathbb{R}$ represents a random variable describing a risk factor dependent on a set of covariates represented by the random vector $X\in\mathcal{X}\subset\mathbb{R}^p$, the objective is to estimate the extreme conditional quantile defined by:
\begin{equation}\label{quantile_inf}
\mathcal{Q}_{x}(\tau) = \inf \{ y : F_{Y|X=x}(y) \geq \tau \},\text{   	    $x \in \mathcal{X}, $}
\end{equation}
with $\tau$ close to $1$ and $F_{Y|X=x}$ being the conditional distribution of $Y|X=x$ (see \cite{koenker_regression_1978}). Observations in the tails of the distribution are often rare and pose a particular challenge when estimating the quantity (\ref{quantile_inf}). This difficulty arises from the limited nature of available data in these extreme regions of the distribution. Indeed, rare events are often crucial in various fields such as risk management, strategic planning, and decision-making. In the context of extreme quantile regression, it becomes imperative to develop robust and accurate methods to handle these infrequent observations, in order to improve the reliability of models and predictions in situations where extreme outcomes play a significant role. This is the case, for example, in risk analysis, in hydrology when estimating a water level that has not yet been observed (\cite{gumbel_statistical_1963}, \cite{hu_evaluation_2020}).

Let \( n \) denote the sample size available for analysis and \( \tau = \tau_n \) (dependent on the sample size) the level of the quantile we seek to estimate. Classical quantile estimation methods work well when \( n(1-\tau_n) \to \infty \) as \( \tau_n \to 1 \) (for \( n \to +\infty \)). In this case, the quantile we seek to estimate is within the sample, and there is also a large amount of data in the region of the quantile to be estimated.
However, the situation differs when  $n (1-\tau_n) \in [0, +\infty[$ as $\tau_n \to 1$ (when $ n \to +\infty$). In this latter case, estimation involves extrapolating beyond the data range or into the distribution's tail, and $\tau_n$ is referred to as an extreme quantile level. In other words, the sought quantile lies outside the range of the available sample, making estimation more complex and requiring specific approaches to handle these borderline situation. Thanks to the asymptotic results of extreme value theory (\cite{de_haan_extreme_2006}), extrapolation beyond this range is possible. Based on this theory, \cite{chernozhukov_extremal_2005} proposed an extreme quantile regression method within the framework of a linear relationship between the response variable  $Y$ and the predictor variable  $X \in \mathbb{R}^p$. Several other authors have proposed extreme quantile regression models, including \cite{yu_bayesian_2001}, \cite{koenker_quantile_2001}, \cite{daouia_kernel_2013}, \cite{takeuchi_nonparametric_2006}, \cite{laksaci_estimation_2009}, \cite{gardes_estimating_2015}, \cite{gardes_integrated_2019}, and \cite{el_methni_extreme_2017}.

The second difficulty faced by classical extreme quantile regression methods arises when the dimension of the predictor space $\mathcal{X}$ is large (i.e., $p$ is large) and the relationship between the characteristic variables and the response variable is complex. In high dimensions, a simple quantile regression model may introduce additional bias (\cite{gnecco_extremal_2024}). Recent literature has seen the development of machine learning approaches to address this challenge. First, we can cite the work of \cite{meinshausen_quantile_2006}, who proposed a method based on the random forest method of \cite{breiman_random_2001}. \cite{athey_generalized_2019} proposed a quantile regression model based on the generalized random forest method. Their methods encounter difficulties mainly when the quantile of interest is extreme (as shown by the simulation studies in section \ref{sect_Application}). 
Quantile regression methods that combine extreme value theory and machine learning have emerged very recently and are competitive with existing methods when the quantile of interest is extreme and when the predictor space is high-dimensional. To our knowledge, the first model proposed in this sense is that of \cite{farkas_generalized_2024}, based on regression trees and the Peaks Over Threshold (POT) approach of extreme value theory. Next, we can mention the method of \cite{velthoen_gradient_2023}, which models the distribution of equation (\ref{quantile_inf}) using the conditional generalized Pareto distribution and the gradient boosting method. \cite{pasche_neural_2023} propose a method that still combines extreme value theory to approximate the conditional distribution of equation (\ref{quantile_inf}) and neural networks to estimate the parameters of this distribution. \cite{gnecco_extremal_2024} propose a method combining the POT approach and generalized random forests.

Previous work has mainly relied on the \textit{Peaks Over Threshold}  approach to address the problems associated with estimating extremes. Although this method is widely recognized for its effectiveness, it nevertheless presents several notable limitations, particularly when the only available data consist of block maxima aggregated over fixed periods — such as annual maxima derived from long historical time series or large-scale simulations. In such contexts, the block maxima (BM) approach offers a more suitable alternative. This method is extensively used in environmental sciences, where the Generalized Extreme Value (GEV) distribution has proven particularly well-suited for modeling phenomena such as annual or monthly maximum temperatures, or extreme river discharges. As highlighted by \cite{ferreira_block_2015}, the BM approach presents several practical advantages. It is particularly appropriate when only periodic maxima data are available. Moreover, it allows for a certain degree of temporal dependence within blocks — for instance, in the presence of seasonality or short-term dependence — which is advantageous in situations where the assumption of independent and identically distributed observations is difficult to justify. Finally, the BM approach is often simpler to implement, as block periods naturally align with existing temporal structures (e.g., years or seasons). In this work, we focus on addressing the problem of extreme quantile regression using the block maxima approach. This method constitutes a relevant alternative to POT-based method in many applied contexts, particularly when data availability or temporal structure constraints hinder the use of threshold exceedance methods.
The generalized extreme value  distribution is given by:
\begin{equation}\label{gev_dist}
G_{(\mu,\sigma,\xi)}(z)=\begin{cases}
\exp\left(-(1+\xi \dfrac{z-\mu}{\sigma})_{+}^{-\frac{1}{\xi}}\right)&  \text{ if} \quad\xi \neq0,\\
\exp \left(-\exp (-\frac{z-\mu}{\sigma})\right)& \text{ if} \quad \xi=0, 
\end{cases}
\end{equation}
defined on $\{z: 1+\xi \dfrac{z-\mu}{\sigma}>0 \}$ 
where \(\mu \in \mathbb{R}\), \(\sigma > 0\), and \(\xi \in \mathbb{R}\) are respectively the location, scale, and shape parameters, with \(a_{+} = \max\{0, a\},$ $\forall a \in \mathbb{R}\).

In this study, applying the block maxima approach, we approximate the conditional distribution \(Y|X=x\) of block maxima by a generalized extreme value distribution, whose parameters are functions of the features \(x\in \mathcal{X}\). These parameters are estimated by a weighted maximum likelihood method, with weights obtained using the generalized random forest method developed by \cite{athey_generalized_2019}. Section~\ref{sect_Background} provides a reviews of quantile regression, the BM approach to extreme value theory and the generalized random forest method. Section~\ref{sect_model} presents our proposed model in detail. Section~\ref{sect_Application} examines the results obtained when applying our extreme quantile regression method to simulated data under several scenarios.  Finally, section \ref{section_real_appli} presents an application to daily weather data from the Fort Collins station in Colorado (USA).

	\section{Background}\label{sect_Background}	
\subsection{Generalized extreme value distribution}\label{sect_GEV}
In this section, we will present the approach of extreme value theory that we will use to address the first quantile regression challenge mentioned in the introduction: the block maxima approach. This method is based on the results of \cite{fisher_limiting_1928} and \cite{gnedenko_sur_1943}, which provide the possible asymptotic distributions of the maximum of a sequence of random variables $X_1, \cdots, X_m$ drawn independently and identically distributed from a variable $X$ with probability distribution $F$. These authors showed that there exist normalization constants $a_m>0$ and $b_m \in \mathbb{R}$ such that:
\begin{equation}\label{dom_attraction}
\lim_{m\to +\infty} F^{m}(a_mx+b_m) =G_{\xi}(x), \text{   $x\in \mathbb{R}$ and $\xi \in \mathbb{R},$ }	
\end{equation} 
where $G_{\xi}$ is a non-degenerate probability distribution defined by:	$$	G_{\xi}(x) = \exp\left(-(1+\xi x)^{-\frac{1}{\xi}}\right) \text{ with } 1+\xi x > 0,$$
which is called the extreme value distribution. Any function $F$ that satisfies equation (\ref{dom_attraction}) is said to belong to the max-domain of attraction of the extreme value distribution $G_{\xi}$ and is denoted as $F \in \mathcal{D}(G_{\xi})$ (\cite{de_haan_extreme_2006}).  Considering \(Y_1, \cdots, Y_N\) as a sequence of independent and identically distributed random variables according to the random variable \(Y\) with distribution function \(F\), the block maxima  method involves dividing the data into \(n\) blocks of equal size \(m > 1\) (or nearly equal), denoted as $$B_{k,m}=\{Y_{(k-1)m+1},\cdots,Y_{km}\}, \text{    with $k=1,\cdots,n$.}$$ For any \(m > 1\), the distribution of \(Z_k=\max_{B_{k,m}}{(Y_{i})}\) is \(F^m\) and satisfies (\ref{dom_attraction}) with a certain normalization constant \(a_m>0\) and \(b_m \in \mathbb{R}\). Therefore, its distribution is given by the generalized extreme value  distribution with parameters \((a_m, b_m, \xi_0)\). The BM method assumes that these block maxima, denoted \(Z_{k}\), exactly follow the GEV distribution (explicitly defined by (\ref{gev_dist})), and that the sequence of random variables \(Z_1, \cdots, Z_n\) thus formed is also independent and identically distributed. The specificity of this method lies in the choice of the optimal block size. Increasing the block size \(m\) leads to an increase in the variance of the estimation, while decreasing the block size introduces a bias. Therefore, it is essential to find a trade-off between bias and variance when defining the blocks to ensure accurate estimation. The BM method is commonly presented, discussed, and employed in the literature for modeling extremes. Among the reference works, one can refer to \cite{coles_introduction_2001-1} and \cite{de_haan_extreme_2006}.
The general form of the Generalized Extreme Value  distribution, introduced by \cite{jenkinson_frequency_1955}, is given by the function \( G_{(\mu, \sigma, \xi)} (.) \) defined in equation~(\ref{gev_dist}).
Depending on the sign of $\xi,$ three domains of attraction are distinguished: the Fréchet domain of attraction if $\xi>0,$ the Gumbel domain of attraction if $\xi=0,$ and the Weibull domain of attraction if $\xi<0.$ The $\tau-th$ extreme quantile  of the Generalized Extreme Value  distribution is obtained using equation (\ref{quantile_inf})  and is given by:
\begin{equation*}\label{GEVquantile}
Q_{\tau}= \begin{cases}
\mu + \dfrac{\sigma }{\xi}\left(\left(-\ln(\tau)\right)^{-\xi }-1\right)  & \text{if  }  \xi \neq 0,\\
\mu - \sigma
\ln \left( -\ln\left(\tau\right)\right)  & \text{if  } \xi= 0,
\end{cases}\end{equation*}
where $\tau$ close to 1, \(\mu \in \mathbb{R}\), \(\sigma > 0\), and \(\xi \in \mathbb{R}\). Estimating the quantile of the GEV distribution therefore requires the estimation of the parameters $\mu$, $\sigma$ and $\xi$. Among the various methods available, the maximum likelihood method is the most commonly used and is the focus of this study. The theoretical validity of this method for GEV parameter estimation has been demonstrated by \cite{dombry_existence_2015,dombry_maximum_2019} and \cite{bucher_maximum_2017}. Denoting the parameter vector of GEV as $\theta=\left(\mu,\sigma,\xi\right),$ the negative log-likelihood of the GEV distribution for the sample $z_1,\cdots,z_n$ is given by:
\begin{equation}\label{log_vraissemblance}
L(\theta;z_1,\cdots,z_n)=\dfrac{1}{n}\sum_{i=1}^{n}\ell_{\theta}(z_i),
\end{equation}
where $\ell_{\theta}$ is defined by:
$$\ell_{\theta}(z_i)=  \begin{cases}
log(\sigma)+(1+\frac{1}{\xi})\log \left(1+\xi\frac{z_i-\mu}{\sigma}\right)+\left(1+\xi\frac{z_i-\mu}{\sigma}\right)^{-\frac{1}{\xi}} & \text{ if   $ 1+\xi\frac{z_i-\mu}{\sigma}>0,$ }\\
+\infty & \text{otherwise},
\end{cases}$$  when  $\xi \neq 0 ,$ and

	$$\ell_{\theta}(z_i)= \log(\sigma)+\left(\frac{z_i-\mu}{\sigma}\right)+\exp{\left(-\frac{z_i-\mu}{\sigma}\right)},$$  when   $\xi= 0.$ 
The maximum likelihood estimator $\hat{\theta}$ satisfies \[\hat{\theta} \in \arg\min_{\theta \in \Theta}L(\theta ; z_1,\cdots,z_n),\]
where $\Theta \subset \mathbb{R}\times]0,+\infty[\times\mathbb{R}.$ 
\subsection{Generalized random forests}\label{sect_grf}
In this section, we introduce the machine learning method we will be using to address the second problem, which aims to handle the high-dimensional space $\mathcal{X}$ of predictors and capture the complex relationship between the response variable and the features. The generalized random forests (\cite{athey_generalized_2019}) is a generalization of the classical random forest method proposed by \cite{breiman_random_2001}. It retains the attractive features of classical random forests but allows customization of the loss function used for tree construction. Random forest is a learning method used for both classification and regression tasks. It belongs to the family of ensemble learning methods and provides a non-parametric estimation of the conditional mean. It involves aggregating $B$ trees trained in parallel on bootstrap samples from the original training set, similar to the bagging method introduced by \cite{breiman_bagging_1996}.
A key difference with bagging lies in how feature variables are used to split nodes in each tree. Instead of using all features at every node split, a random subset of features is selected. This random selection introduces additional variability between the trees, promoting diversity and improving the model's generalization (\cite{breiman_random_2001}).
Each tree provides an estimate of the conditional mean, which is obtained as the function minimizing the mean squared error. If $(X_i,Y_i) \in \mathcal{X} \times \mathcal{Y}$ where $\mathcal{X} \subset \mathbb{R}^p$ and $\mathcal{Y} \subset \mathbb{R}$, random forests are used to estimate $\mu(x) = \mathbb{E}(Y_i | X_i = x).$
Let $\eta_b(x)$ denote the predicted value by the $b^{th}$ tree for data $x \in \mathcal{X}$. In the case of regression, this prediction is given by:
\[
\eta_b(x)=\sum_{i=1}^{n}\dfrac{\mathbb{1}_{\{X_i\in R_b(x)\}Y_i}}{|\{i:X_i\in R_b(x)\}|}, \text{  } b=1,\cdots,B,
\]
where $R_b(x) \subset \mathcal{X}$ denotes the region containing $x$ in tree $b$, and $|E|$ denotes the cardinal  of $E$. The random forest gives as its prediction the average of the predictions of the $B$ trees, thus it predicts the value $\eta(x)$ given below for data $x$:
\begin{eqnarray} \notag
\eta(x)&=&\dfrac{1}{B}\sum_{b=1}^{B}\eta_b(x)\\ \notag
&=& \sum_{i=1}^{n}w_{n}(x,X_i)Y_i,
\end{eqnarray}
with 
\begin{equation}\label{forests_weights}
w_{n}(x,X_i)=\dfrac{1}{B}\sum_{b=1}^{B}\dfrac{\mathbb{1}_{\{X_i\in R_b(x)\}}}{|\{i:X_i\in R_b(x)\}|}.
\end{equation}
The $w_{n}(x,X_i)$, for $i=1,\cdots,n$, represent the similarity weights assigned by the random forest for each observation $X_i$ given $x \in \mathcal{X}$.

 While the similarity weights $w_n(x, X_i)$ assigned by the generalized random forests (GRF) are computed using the same formula as in the classical forest, the key difference lies in the loss functions employed during tree construction. In classical random forests, a high weight is typically assigned to an observation $X_i$ when $\mathbb{E}(Y | X = X_i) \approx \mathbb{E}(Y | X = x)$ (\cite{meinshausen_quantile_2006}). However, as shown by \cite{athey_generalized_2019} (Figure 2) and \cite{gnecco_extremal_2024} (Figure 1), these weights can still be large even when $\mathcal{Q}(Y | X = X_i) \not\approx \mathcal{Q}(Y | X = x)$, indicating that classical forests often fail to capture heterogeneity in the conditional quantile function. This limitation is addressed in GRF by incorporating the quantile loss into the tree construction process, enabling the method to target conditional quantiles directly. This approach has been successfully applied for both moderate and extreme quantile modeling by \cite{athey_generalized_2019} and \cite{gnecco_extremal_2024}, respectively.
Forests-based methods have the advantage of requiring few tuning parameters to provide effective predictions. This characteristic makes generalized random forests  a natural choice for constructing similarity weights, especially as they capture the heterogeneity of conditional quantiles, a central aspect in our study. Unlike classical approaches such as kernel or nearest-neighbor methods, which are often limited when faced with non-linear structures or high-dimensional data, GRF exploit tree structure to accurately model local variability. They thus produce similarity weights that are both more robust and better suited to estimating extreme quantiles. What's more, compared with techniques such as gradient boosting or neural networks, GRF combine simplicity of use with sound theoretical foundations (\cite{athey_generalized_2019}). They also offer better high-dimensional adaptability than generalized additive models (\cite{koenker_additive_2011}). Furthermore, the BM approach is better suited to random forests, as it generates i.i.d. observations from fixed block maxima, in line with the assumptions underlying random forests. Conversely, the POT method produces potentially correlated excesses, requiring additional adjustments, and relies on an often empirical threshold, which can weaken learning. This combination of robustness and flexibility makes GRF-based weights particularly well-suited to our framework. For clarity, we denote $w_i(x) \equiv w_n(x, X_i)$ for all $x \in \mathcal{X}$ and $i = 1, \ldots, n$.

\subsection{ Quantile regression }\label{sect_quantile}
In classical regression, given a random vector $(X,Y)$, we seek to predict the conditional mean $E[Y|X=x]$, which is a quantity that summarizes the behavior of $Y$ given $X=x$ around the  center of distribution. However, applications in risk assessment require knowledge of the tail behavior of its distribution. Quantile regression, introduced by \cite{koenker_regression_1978}, addresses this by providing a richer description than classical regression. The latter focuses on the estimation of the conditional quantile defined in (\ref{quantile_inf}), which in the case of a continuous conditional distribution is given by:
\begin{equation}\label{quantile_inverse}
\mathcal{Q}_{\tau}(Y|X=x)= F^{-1}_{Y|X=x}(\tau), \text{  $ x \in \mathcal{X}$ and  $\tau \in (0,1).$}
\end{equation}
In the following, we denote $\mathcal{Q}_{x}(\tau)$ to represent the conditional quantile function $\mathcal{Q}_{\tau}(Y|X=x)$. To the best of our knowledge, the first appearance of the random forest method in quantile regression, addressing our second problem stated in the introduction, dates from  the work of \cite{meinshausen_quantile_2006}. Considering $(X_1,Y_1), \ldots, (X_n,Y_n)$ as a sequence of independent and identically distributed random vectors from $(X,Y) \in \mathcal{X} \times \mathcal{Y}$, the author estimates the conditional distribution $F(.|X=x)$ by:
\begin{equation*}
\hat{F}\left(y|X=x\right)=\sum_{i=1}^{n} w_n(x,X_i) \mathbb{1}_{\{Y_i\leq y\}}, \text{   $y\in \mathbb{R}$ and $x \in \mathcal{X},$}
\end{equation*}
where the weights $w_n(x,X_i)$ are obtained using the random forest method by \cite{breiman_random_2001} as defined in (\ref{forests_weights}). Another method of quantile regression is proposed by \cite{athey_generalized_2019} using the generalized random forest method described in section (\ref{sect_grf}), with the quantile loss function $$\rho_{\tau}(c) = c(\tau - \mathbb{1}_{\{c<0\}}).$$ 
Although these methods address high-dimensional problems and capture complex relationships between the dependent variable and covariates, they encounter difficulties when  the order of the  quantile is extreme. Additional research has therefore been conducted to address this issue by combining extreme value theory, mainly the POT approach, and statistical learning methods. Among these efforts, notable contributions include \cite{farkas_generalized_2024}, \cite{gnecco_extremal_2024}, \cite{velthoen_gradient_2023}, and \cite{pasche_neural_2023}. The quantile regression method we propose and detail in the next section builds upon these recent developments.

\section{ GEV Extremal Random Forest }\label{sect_model}

In this section, we present our  method of extreme quantile regression that addresses the issues outlined in the introduction. To address the first issue, we model the tail of the conditional distribution $F(.|X=x)$ from equation (\ref{quantile_inverse}) using the generalized extreme value distribution, as described in section (\ref{sect_GEV}). To tackle the second issue, we employ the generalized random forest method to determine the weights $w_n(x,X_i)$, which will be used to estimate the parameters of the GEV distribution. These parameters are estimated using weighted maximum likelihood estimators, following the approach proposed by \cite{gnecco_extremal_2024} within the peaks-over-threshold  framework.
To proceed, let's consider an independent and identically distributed sample $(X_1,Y_1), \ldots, (X_N,Y_N)$ from the random vector $(X,Y)$. Assume  $N = n \times m$ and divide the sample into $n$ blocks of size $m$, such that the $k^{th}$ block for $k=1,\ldots,n$ is given by:
\begin{equation*}
B_{k,m}=\{(X_{(k-1)m+1},Y_{(k-1)m+1}),\cdots,(X_{km},Y_{km})\}.
\end{equation*}
Noting 
 $Z_{k,m}=\max \{ Y_i: (X_i,Y_i)\in B_{k,m}\}$
and  $X_{k,m}$ the $X_i$ corresponding to the $Y_i$ maximizing  $\{ Y_i: (X_i,Y_i)\in B_{k,m}\}$.
Thus, for all $m>1$, $\{(X_{k,m}, Z_{k,m})\}_{k=1,\ldots,n}$ represents the sample of block maxima relative to the variable $Y$. For any $x \in \mathcal{X}$, we assume that the distribution $F_x$ of $Z_{k,m} | X_{k,m}=x$ follows a Generalized Extreme Value  distribution with  parameter depending to the covariate $x.$ More precisely, 
$\mu(\cdot): \mathcal{X} \to \mathbb{R}$ is the location parameter function,
$\sigma(\cdot): \mathcal{X} \to \mathbb{R}_+^{*}$ is the scale parameter function and $\xi(\cdot): \mathcal{X} \to \mathbb{R}$ is the shape parameter function.
The conditional GEV distribution is obtained by substituting $\theta = (\mu, \sigma, \xi)$ in (\ref{gev_dist}) with $\theta(x) = (\mu(x), \sigma(x), \xi(x))$ for all $x \in \mathcal{X}$. The $\tau$-th extreme quantile  of the conditional GEV is obtained using equation (\ref{quantile_inverse}) and is given for all $x \in \mathcal{X}$ by:
\begin{equation}\label{GEVquantile}
Q_{x}(\tau)= \begin{cases}
\mu(x) + \dfrac{\sigma (x)}{\xi (x)}\left(\left(-\ln(\tau)\right)^{-\xi (x)}-1\right)  & \text{if  }  \xi (x)\neq 0,\\
\mu(x) - \sigma(x)\ln \left( -\ln\left(\tau\right)\right)  & \text{if  } \xi(x)= 0 .
\end{cases}\end{equation}
Estimating this conditional quantile involves estimating the parameters $\mu(x)$, $\sigma(x)$, and $\xi(x)$ for all $x \in \mathcal{X}$. Our proposed method here consists of estimating these parameters and then substituting them into equation (\ref{GEVquantile}) to obtain the estimation of $Q_{x}(\tau)$.
For this purpose, we propose a weighted form of the maximum likelihood estimator given in (\ref{log_vraissemblance}). As proposed by \cite{gnecco_extremal_2024} for the POT approach, we estimate $\theta(x)$ by $\hat{\theta}(x) = (\hat{\mu}(x), \hat{\sigma}(x), \hat{\xi}(x))$, which is a weighted maximum likelihood estimator, i.e., it minimizes $L_n(\theta;x)$ defined by:

\begin{equation}\label{weigth_log_likelihood}
L_n(\theta;x)=\sum_{i=1}^{n}w_i(x)\ell_{\theta}(z_i) \text{  for all }	 x\in \mathcal{X},
\end{equation}
with $\ell_{\theta}(z_i)$ defined in section (\ref{sect_GEV}).
To capture the heterogeneity of the quantile function and improve estimation in a high-dimensional feature space, we obtain the weights $w_i(x)$ for $i=1, \ldots, n$ using the generalized random forests method, as described in section (\ref{sect_grf}).

It should be noted that the support of the GEV distribution depends on the extreme value index $\xi$, which is unknown (see \cite{dombry_existence_2015}). The usual regularity conditions ensuring good asymptotic properties of the maximum likelihood estimator (MLE) are not satisfied. This problem is first studied by \cite{smith_maximum_1985}, who shows that there is no maximum likelihood estimator  if $\xi \leq -1$, that asymptotic normality holds if $\xi > -0.5$, and consistency holds if $\xi > -1$. \cite{dombry_existence_2015} demonstrated the existence and consistency of this estimator for the GEV distribution parameters when $\xi > -1$. The asymptotic normality of the MLE for the block maxima approach was proven by \cite{dombry_maximum_2019}. In general, $L_n(\theta; x)$ does not admit a global minimum, and we refer to the work of \cite{dombry_existence_2015} to state that $\hat{\theta}(x)$ is a weighted maximum likelihood estimator if $\hat{\theta}(x)$ is a local minimum. Thus, we define $\hat{\theta}(x)$ to be the vector that minimizes $L_n(\theta; x)$ over a large compact set $\varTheta \subset \mathbb{R} \times ]0, +\infty[ \times ]-1, +\infty[$. The weighted maximum likelihood estimator is given by:
\begin{equation}\label{MLE}
\hat{\theta}(x)\in \arg\min_{\theta \in \Theta} L_n(\theta;x).
\end{equation}
We take the intermediate order $\tau_0 = 0.8$ as considered in the works of \cite{gnecco_extremal_2024} and \cite{velthoen_gradient_2023}. Thus, we estimate the conditional quantile of order $\tau > \tau_0$ by:
\begin{equation}
\hat{Q}_{x}(\tau)= 
\begin{cases}
\hat{\mu}(x) + \dfrac{\hat{\sigma}(x)}{\hat{\xi}(x)} \left(\left(-\ln(\tau)\right)^{-\hat{\xi}(x)} - 1\right) & \text{if } \hat{\xi}(x) \neq 0,\\
\hat{\mu}(x) - \hat{\sigma}(x) \ln\left(-\ln(\tau)\right) & \text{if } \hat{\xi}(x) = 0,
\end{cases}
\end{equation} 
where $\hat{\mu}(x),$ $\hat{\sigma}(x)$ and $\hat{\xi}(x)$ are the estimated parameters obtained using the weighted maximum likelihood method.
The algorithm (\ref{algo_GEV_erf}) details our method for predicting  the extreme conditional quantiles, which we have named GEV-erf. This method combines the use of the generalized extreme value  distribution with generalized random forests to estimate extreme conditional quantiles precisely and efficiently.

Among the parameters of the GEV distribution, the shape parameter $\xi$ is of particular importance, as it provides information on the shape of the tail of the distribution. A distribution is considered to have a heavy tail when $\xi > 0$, a light tail when $\xi = 0$, and a finite endpoint distribution (i.e. $x_F = \{y \in \mathbb{R}, F(y) < 1\}$ finite) when $\xi <0$. Because of its importance, this parameter has received considerable attention in the literature. Many methods have been proposed for estimating $\xi$, the most notable examples being Hill's estimator, which focuses on the most important data points for estimating tail behavior, and Pickands' estimator.
In line with the proposals of \cite{gnecco_extremal_2024} for maximum likelihood in the context of the POT approach, and \cite{bucher_penalized_2021} for the maximum likelihood of the GEV, we penalize the negative log-likelihood in the  following form:

\begin{equation*}\label{weigth_log_likelihood}
L^{pen}_n(\theta,x)=\sum_{i=1}^{n}w_i(x)\ell_{\theta}(z_i)+ \lambda \left(\xi-\xi_0\right)^2.
\end{equation*}\\
The parameter $\xi_0$ used here corresponds to the shape parameter derived from the unconditional GEV distribution, which means it is obtained by setting $w_i(x)=1$ for all indices $i \in \{1,\cdots,n\}$ in equation (\ref{MLE}). The penalty parameter $\lambda$ is determined by the cross-validation method, thus regularizing the negative log-likelihood in the estimation process. The resulting estimator, $\hat{\theta}_{pen}(x)$, is the weighted and penalized maximum likelihood estimator, which adjusts the GEV parameters by considering both the data and the model complexity, thereby ensuring better generalization. It is defined as :
\begin{equation*}
\hat{\theta}_{pen}(x) \in \arg\min_{\theta \in \Theta} L^{pen}_n(\theta,x).
\end{equation*}
The algorithm to obtain the conditional quantile using $\hat{\theta}_{pen}(x)$ is similar to the algorithm (\ref{algo_GEV_erf}) with the difference that the log-likelihood $L_n(\theta, x)$ is replaced by the penalized likelihood $L^{pen}_n(\theta, x).$  In the simulation study described in section (\ref{sect_Application}), we provide a detailed explanation of the validation method used to determine the parameter $\lambda$. This method includes a series of tests and analyses to ensure the accuracy and reliability of the estimates. Specifically, we discuss the different steps of the validation process, the criteria used to evaluate the performance of the method, and the adjustments made to optimize the estimation of the parameter $\lambda$. The algorithm detailed below outlines our proposed method.
\subsection*{Algorithm}

The algorithm for our extreme quantile regression method, denoted GEV-erf, consists of two sub-algorithms: GEV-erf-fit and GEV-erf-predict. The first sub-algorithm, GEV-erf-fit, is used for training the generalized random forest to obtain the weights $w_i(x)$ and to obtain the sample formed by the block maxima. The second sub-algorithm, GEV-erf-predict, is used to make predictions of the GEV distribution parameters and hence extreme conditional quantiles. It is presented as follows : 
\begin{algorithm}[h!]
\caption{ GEV-erf}
	\label{algo_GEV_erf}
	Let $\mathcal{D}_N=\{(X_i,Y_i)\}_{i=1\cdots N}$ denote the original sample, $\mathcal{D'}_n=\{(X_i,Z_i)\}_{i=1\cdots n}$ denote the sample of block maxima, where $m$ is the block size, and $\alpha$ is the vector containing hyperparameters associated with the generalized random forest.\\
	\begin{algorithmic}[1]
		\BEGIN  $ $ \textbf{GEV-erf-fit}\vspace{0.3cm}
		\Input $ (\mathcal{D}_N$, $m$, $\alpha$) \vspace{0.3cm}
		\STATE   $\mathcal{D'}_n\leftarrow makeBloc(\mathcal{D}_N,m)$\vspace{0.3cm}
		\STATE $w_i(.)\leftarrow GRF(\mathcal{D'}_n,\alpha)$\vspace{0.3cm}
		\Output  GEV-erf $\leftarrow (\mathcal{D'}_n,w_i(.),m)$\vspace{0.3cm}	
	\end{algorithmic}	
	\begin{algorithmic}[1]
		\BEGIN $ $ \textbf{GEV-erf-predict}	\vspace{0.3cm}
		\Input  (GEV-erf, x, $\tau$)\vspace{0.3cm}
		\STATE   $\hat{\theta}(x)\leftarrow \arg\min\limits_{\theta \in \Theta} L_n{(\theta;x)}$\vspace{0.3cm}
		\STATE $\hat{Q}_{x}(\tau)\leftarrow GEV(\hat{\theta}(x))$\vspace{0.3cm}
		\Output $\hat{Q}_{x}(\tau)$ 			
	\end{algorithmic}
	\begin{description}
		\item[$\blacktriangleright$] The function \textbf{makeBloc} divides the initial sample into blocks of a given size \( m \) and returns the sample of block maxima.
		\item[$\blacktriangleright$] The \textbf{GRF} function is used to adjust the weights using the generalized random forests method proposed by \cite{athey_generalized_2019}.		
	\end{description}
\end{algorithm}

\section{Simulation}\label{sect_Application}

In this section, we present a simulation study to demonstrate the effectiveness of our proposed extreme quantile regression method. We generate an independent and identically distributed sample of size $ N = 90,000 $ from the random variable $ X,$ following a uniform distribution over $ [-1,1]^p$. We assume that the conditional distribution of $Y | X = x$ is of the form $ \gamma(x) T_{\nu(x)} $, where $ T_k $ denotes the Student's distribution with $ k $ degrees of freedom. We conduct various experiments by varying the values of $ \gamma(x)$ and $ \nu(x)$. The choice of block size is a key challenge in the BM approach, especially since, as shown in Appendix \ref{Sensitivity_analy}, the GEV-erf model is sensitive to this parameter. We therefore select a fixed block size of $m = 40$ for all scenarios, as our analysis (see Appendix~\ref{Sensitivity_analy}) indicates that model performance remains stable for block sizes ranging from 25 to 50.

In our first simulation study (scenario 1), we set  the dimension of the feature space to $ p = 40$, and evaluate the performance of our method. We then compare its ability to address the first concern stated in the introduction with other methods based on statistical learning approaches. The \( N \) data points are used for the training process, while the evaluation of the method is conducted on an independent sample from the training set, \( \{(x_i, y_i)\}_{i=1}^{n'} \), where \( x_i \in \mathcal{X} \) for \( i = 1, \ldots, n' \) is generated using the Halton sequence method by \cite{halton_algorithm_1964} with $ n' = 8000.$
In the second simulation study (scenario 2), we demonstrate our method's capability to handle the second concern, namely complex dependence on covariates. For this, we conduct simulations with functions \( \gamma(.) \) and \( \nu(.) \) chosen to introduce complex dependence with the covariates.
In the last simulation study of this section (scenario 3), we show the effectiveness of our method in addressing both concerns simultaneously.

We proceed with comparing the performance of our method with two other random forest-based approaches: Quantile Regression Forests (QRF) by \cite{meinshausen_quantile_2006} and Generalized Random Forests (GRF) by \cite{athey_generalized_2019}. To assess these performances, we first employ the Mean Integrated Squared Error (MISE), which is computed as the average of \( m' \) Integrated Squared Error (ISE) values for the estimated conditional quantile, derived from repeating the training and testing process \( m' \) times. This metric aligns with the approach used by \cite{gnecco_extremal_2024} in his work. Additionally, we evaluate performance using two other metrics: the Mean Absolute Error (MAE) and the Median Absolute Error (MedAE). The ISE on the test sample \( \{(x_i, y_i)\}_{i=1}^{n'} \) is defined as follows:
\begin{equation}\label{ISE}
ISE= \frac{1}{n'}\sum_{i=1}^{n'}\left(\hat{Q}_{x_i}(\tau)-Q_{x_i}(\tau)\right)^2,	
\end{equation}
where $x \mapsto Q_{x}(\tau)$ is the true quantile function, $x \in \mathcal{X}$ and $\tau>0.8.$ 
\subsection{Simulations scenarios}

We demonstrate the capability of our proposed method across three scenarios:
\begin{itemize}
	\item [$\rhd$] \textbf{Scenario 1:} We take 
	\[ \gamma(x) = 1 + \mathbb{1}_{\{x_1 > 0\}} \text{  and }
	 \nu(x) = 4 - (x_1^2 - 2x_{2}^2 + x_{3}^2). \]
	The goal is to assess our method's ability to estimate extreme conditional quantiles in the presence of noise and when $p$ is high.
	
	\item [$\rhd$] \textbf{Scenario 2:} We take 
	\[ \gamma(x) = 2 + \frac{1}{1+\exp{(x_1^2+x_2)}} \text{ and } \nu(x) = 4 - (x_1^2 - 2x_{2}^2 + x_{3}^2). \]
	The objective is to evaluate our method's capability to predict extreme conditional quantiles under complex forms of the quantile function. We set the feature size to $p=40$, where only the first three components contribute to the quantile function formation, and the others are noise.
	
	\item [$\rhd$] \textbf{Scenario 3:} We consider 
	\[ \gamma(x) = 1 + 2\pi\varphi(2x_1, 2x_2)  \text{ and } \nu(x) = 3 + \frac{7}{1 + \exp{(4x_1 + 1.2)}}. \]
	In this study, $\varphi$ represents the density of the centered bivariate normal distribution with unit variance and a correlation coefficient of 0.75. The parameters chosen here are the same as those used in experiment 3 of the simulation conducted by \cite{gnecco_extremal_2024}. The aim of this scenario is to evaluate our method's ability to accurately estimate the conditional quantile in situations where the quantile function is complex, the dimension of feature variables is high, and noise is present.
\end{itemize}
\subsection{Parameters tuning choice}
We obtain the penalty parameter $\lambda$ and the  parameters for generalized random forest (specifically min.node.size and num.trees) using cross-validation. Instead of using the classic form (as explained in Section 7.10 of \cite{trevor_hastie_elements_2017}), we follow the approach used by \cite{gnecco_extremal_2024}, using GEV deviance as the metric. More clearly, if we consider that the training sample we have is $\mathcal{D}_n=\{(x_i,z_i)\}_{i=1,\cdots,n},$ and we wish to perform a K-cross validation, the method involves partitioning the sample into $K$ folds of approximately equal size, denoted as $\mathcal{D}^j$ for $j=1,\cdots,K$. For each tuning parameter $\alpha_1,\cdots,\alpha_S$, the model is trained on $\mathcal{D}_n\backslash \mathcal{D}^j$ using the GEV-erf-fit algorithm, and then parameters $\hat{\theta}(x)$ are estimated on $\mathcal{D}^j$ for each $j=1,\cdots,K$. The performance metric, here the negative log-likelihood of GEV, is calculated for each fold, and the average error across the $K$ folds provides an overall estimate of model performance. The $\alpha_s$ achieving the best performance is selected, corresponding to minimizing the cross-validation error (CV-error) across $s,$
\begin{equation}\label{CV}
CV(\alpha_s)=\frac{1}{K}\sum_{j=1}^{K} \sum_{(x_i,z_i )\in\mathcal{D}^j } \ell_{(\hat{\theta}(x_i),\alpha_s)}(z_i),
\end{equation}
where $\ell_{(\hat{\theta}(x),\alpha_s)}$ represents the negative log-likelihood defined in (\ref{log_vraissemblance}) and the $\alpha_s$ for all $s \in \{1,\cdots,S\}$ designate the tuning parameters from which we seek to select the optimal choice. The selection of the parameters $\lambda$ and $min.node.size$ for the different scenarios is detailed in section (\ref{choix_hyperpara}) of the appendix. Our model is called GEV-erf, but we use the abbreviation "gev" to represent it in the various plots and tables showcasing the model performances.

\subsection{Performance of GEV-erf with Scenario 1}
\begin{figure}[h!]
	\centering
	\includegraphics[width=0.991\linewidth,height=0.4\textheight]{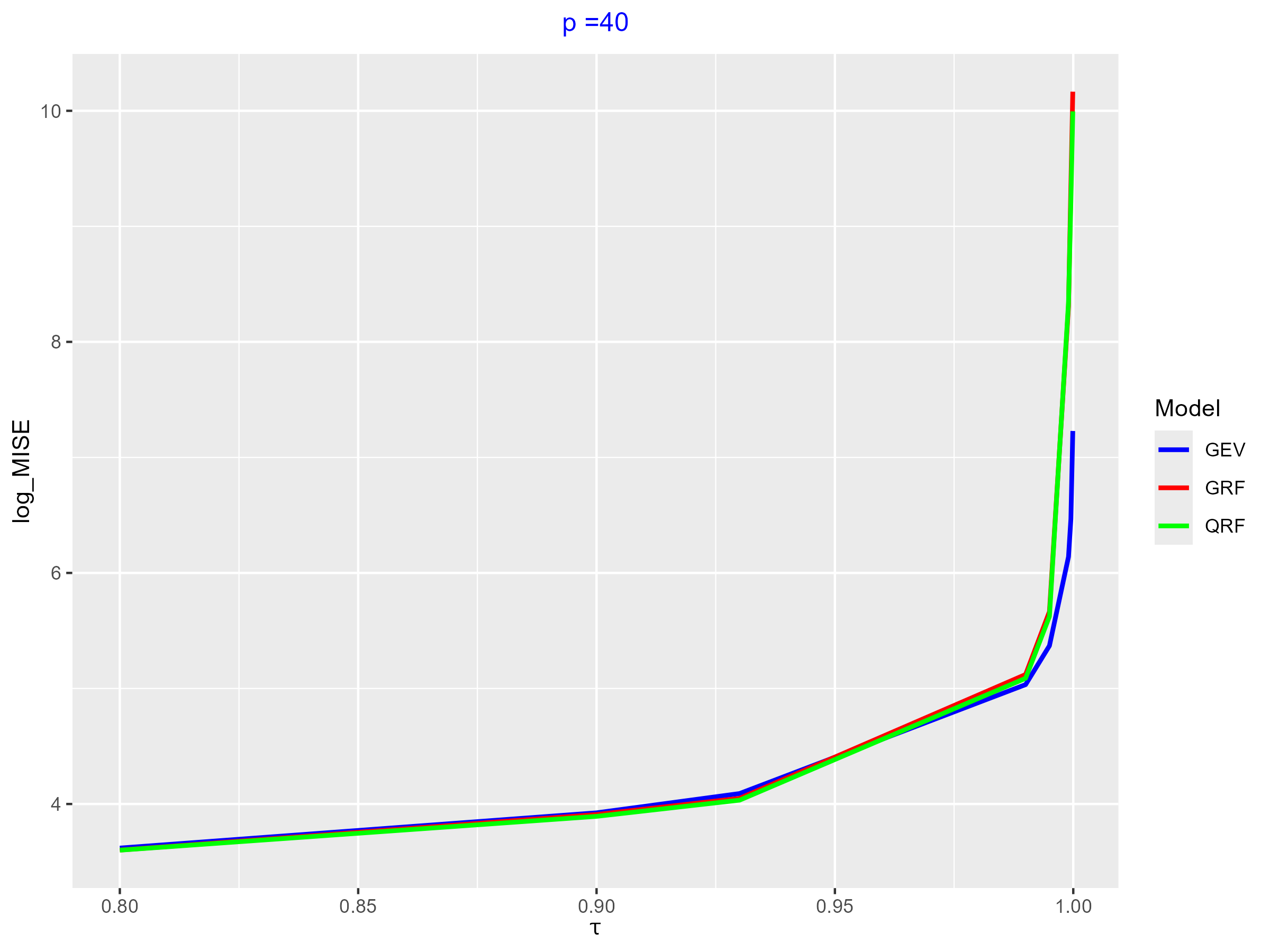}
	\caption{Logarithm of MISE for different methods as a function of $\tau$.}
	\label{fig:experience1quantilevecp401}
\end{figure}

In Scenario 1, we rigorously evaluated the performance of our method by comparing it with Quantile Regression Forest (QRF) proposed by \cite{meinshausen_quantile_2006} and Generalized Random Forest (GRF) by \cite{athey_generalized_2019}. For this comparative analysis, we fixed the number of predictors to $p = 40$ and plotted $\log (\text{MISE})$ against the quantile level $\tau$. The MISE was computed from $m' = 50$ repetitions of the ISE (defined in (\ref{ISE})), ensuring robust and statistically significant evaluation.
Figure~\ref{fig:experience1quantilevecp401} clearly illustrates that our model is competitive compared to QRF and GRF models, especially for moderate quantile orders. This figure also highlights that our method provides better estimation of conditional quantiles, particularly as the quantile level is closed to~1.

Furthermore, our method demonstrates increased stability and reduced variance in quantile predictions, thanks to our innovative approach of local weighting. These results suggest that our model not only competes with existing methods but also outperforms them in scenarios where precise estimates of extreme quantiles are crucial. This is particularly relevant in applications requiring accurate risk management and decision-making based on reliable predictions of rare events.

\subsection{Performance of GEV-erf with Scenario 2}
\begin{figure}[h!]
	\centering
	\includegraphics[width=0.99\linewidth,height=0.5\textheight]{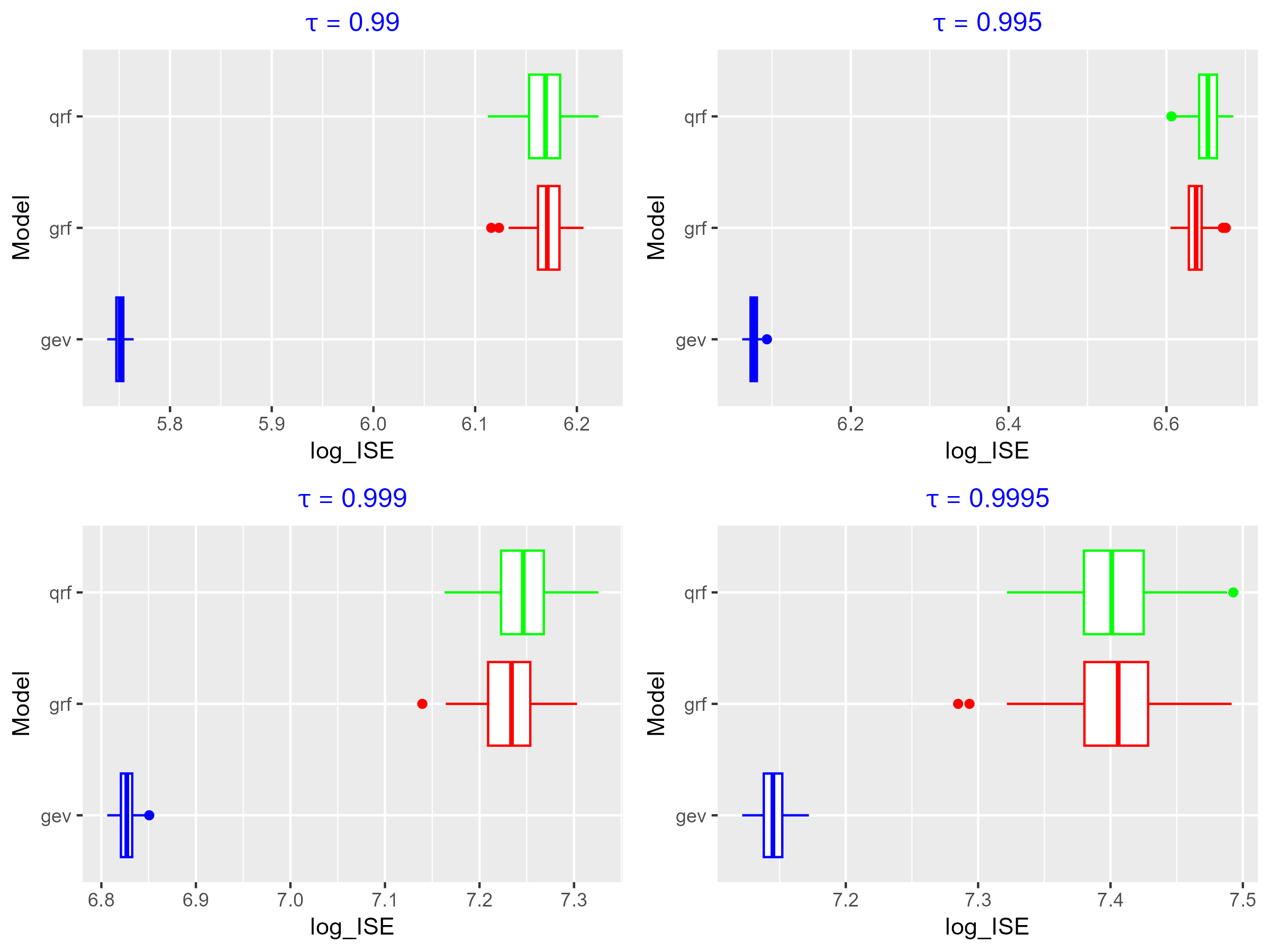}
	\caption{ Boxplot of $log(ISE)$ over $100$ replication, for $p=40$ and $\tau=(0.99,0.995,0.999,0.9995)$}
	\label{fig:scenario22boxplot}
\end{figure}

In the figure~\ref{fig:scenario22boxplot}, we evaluate the performance of our model for scenario 2. To do this, we present box plots of log(ISE) for different quantile orders $\tau~\in~\{0.99, 0.995, 0.999, 0.9995\}$, with $p$ fixed at $40.$ The functions $\gamma(.)$ and $\nu(.)$ are chosen to make the conditional quantile function complex. The boxplots in this figure demonstrate that our model outperforms the QRF model  and the GRF model for various quantile orders. This indicates the ability of our model to make accurate predictions even when the quantile function is complex. 
This scenario showcases our method's ability to effectively predict the conditional quantile of the dependent variable, outperforming the GRF and QRF methods even in the presence of noise and when $\mathcal{Q}_{\tau }(x)$ is complex. This confirms that our method addresses the previously outlined challenges, providing accurate predictions under difficult conditions where the relationship between the conditional quantile function and the covariates is complex and noisy. 

\subsection{Performance of GEV-erf with Scenario 3}
\begin{figure}[h!]
	\centering
	\includegraphics[width=0.99\linewidth,height=0.5\textheight]{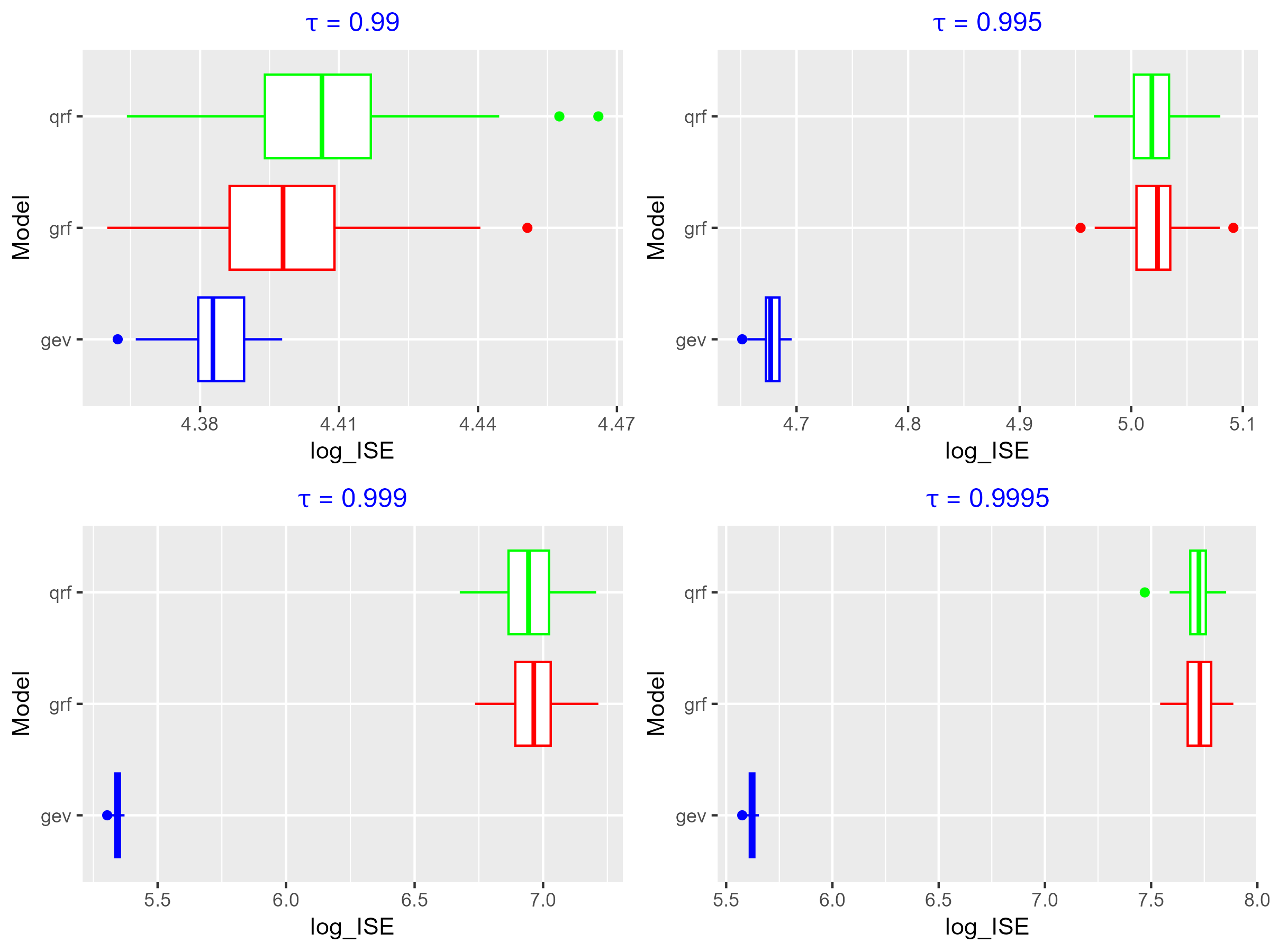}
	\caption{ Boxplot of $log(ISE)$ over $100$ replication, for $p=50$ and $\tau=(0.99,0.995,0.999,0.9995).$}
	\label{fig:scenario32boxplot}
\end{figure}

\begin{figure}[h!]
	\centering
	\includegraphics[width=0.99\linewidth,height=0.5\textheight]{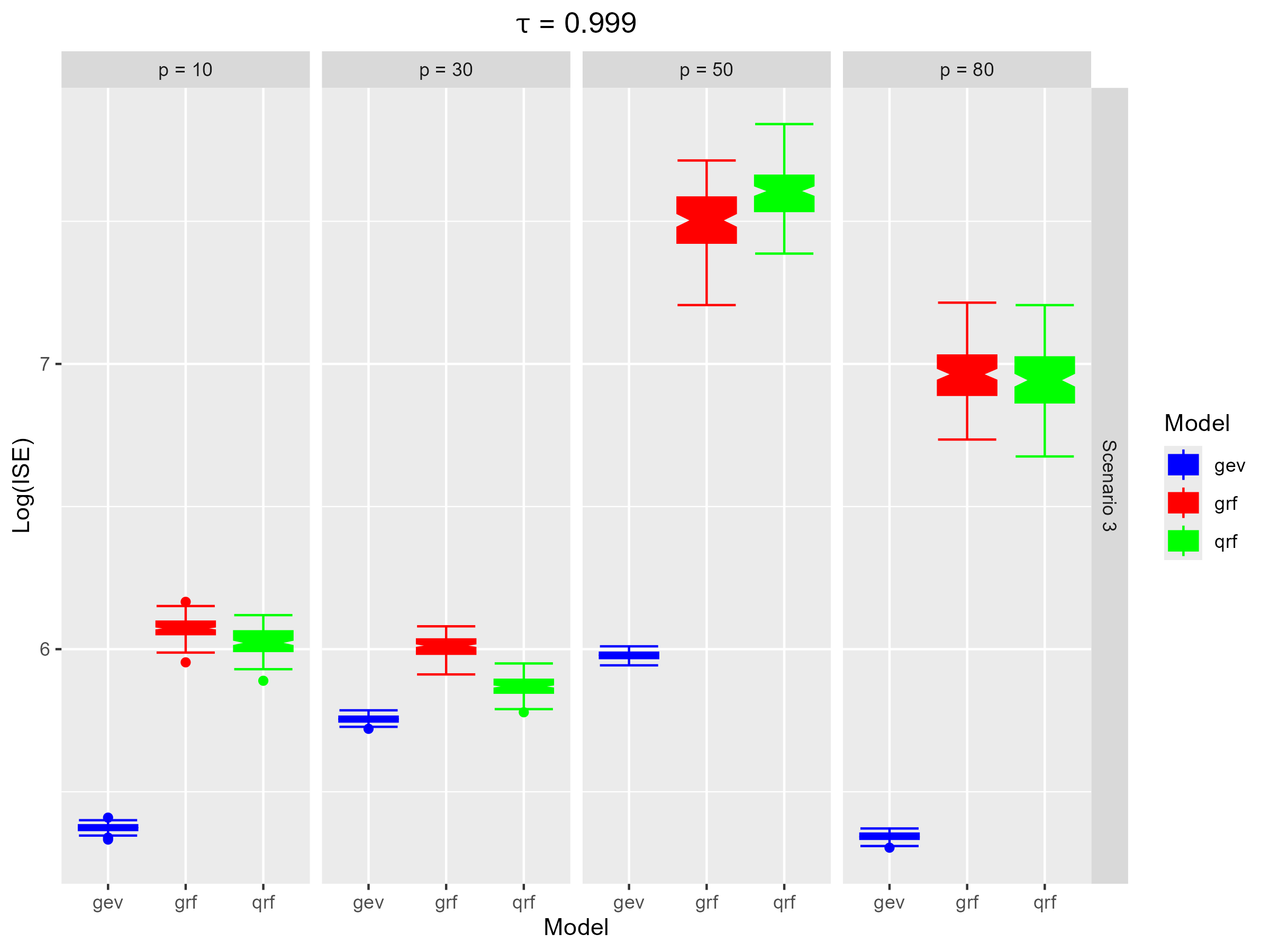}
	\caption[]{\small Boxplot of $log(ISE)$ over $100$ replication, for $p \in \{10,30,50,80\}$ and $\tau=0.999.$}
	\label{fig:boxplotgridscen3}
\end{figure}

In this scenario, the covariate size is fixed at $p=50$, and the functions $\gamma(.)$ and $\nu(.)$ are chosen to make more complex the relationship between the covariates and the conditional quantile of the dependent variable. This scenario highlights the performance of our GEV-erf method in predicting extreme quantiles when the covariate size is large, the quantile function is complex, and in the presence of noise. Figure~\ref{fig:scenario32boxplot} displays the boxplot of $\log(ISE)$ for our model as well as for the GRF and QRF models, across different quantile orders. 
This figure demonstrates that our GEV-erf method outperforms the GRF and QRF methods, indicating its effective capture of the complex structure of extreme quantiles, its adaptation to the high dimensionality of covariates, and its ability to address the challenges outlined in the introduction. In figure~\ref{fig:boxplotgridscen3}, we present the boxplot of $\log(ISE)$ for various methods, considering different covariate sizes and with quantile order set at $0.999$. This graph shows that our method is efficient for different values of $p$.
In terms of performance, our method is superior to GRF and QRF for extreme quantile regression analysis, particularly when the quantile function is complex and the size of the covariates is large. This is particularly evident in scenarios where capturing the subtleties of the quantile function is crucial. The performance of our method remains robust and reliable, highlighting its advantages over other methods. 

We evaluate the performance of the models using two additional error metrics: the mean absolute error (MAE) and the median absolute error (MedAE). The results obtained are summarized in Tables \ref{tab:performanceScena1}, \ref{tab:performanceScena2}, and \ref{tab:performanceScena3}, corresponding to the three scenarios of the study. The analysis of these results, conducted for three levels of high quantiles ($\tau = 0.99, 0.995, 0.9995$), demonstrates that the GEV-erf model systematically outperforms the GRF and QRF models. Indeed, GEV-erf exhibits significantly lower errors as well as increased stability, even for extreme quantiles. Conversely, the GRF and QRF models show a strong sensitivity to the highest quantiles ($\tau = 0.9995$), resulting in a marked increase in errors. Moreover, the detailed analysis by scenario confirms this trend: in scenario 1, GEV-erf maintains stable performance, while the errors for the GRF and QRF models increase sharply for extreme quantiles; in scenario 2, a general degradation in performance is observed, but GEV-erf retains relatively higher accuracy; finally, in scenario 3, overall performance improves for all models, with GEV-erf nonetheless confirming its superiority. In summary, the analysis highlights the robustness and precision of the GEV-erf model for estimating high quantiles, while the GRF and QRF models reveal their limitations in dealing with extreme quantiles. These findings further support the conclusions previously obtained.

\begin{table}[h!]
	\centering
	\renewcommand{\arraystretch}{1.2}
	\begin{tabular}{|c|c|c|c||c|c|c|}
		\hline
		\multicolumn{7}{|c|}{\textbf{Scenario 1}}\\
		\hline
		\multirow{2}{*}{\centering \textbf{Model}}& \multicolumn{3}{c|}{\textbf{MAE}} & \multicolumn{3}{c|}{\textbf{MedAE}} \\ 
		\cmidrule{2-7}
		& $\tau = 0.99$   & $\tau = 0.995$ & $\tau = 0.9995$ & $\tau = 0.99$  & $\tau = 0.995$ & $\tau=0.9995$ \\ \hline
		\textbf{gev}            &  12.4                                     &  14.6                              &  24.9                &  12.3                       & 14.7                             & 25.5                         \\ \hline
		\textbf{grf}            &12.8                        &     16.5                          &   71.8                 & 12.8                                &        16.3                        &   31.7                       \\ \hline
		\textbf{qrf}            & 12.5                                           &  15.9                             &    68.7                &      12.5                              &    15.1                             &   31.1                      \\ \hline	
	\end{tabular}
	\caption{Performance of models for different metrics.}
	\label{tab:performanceScena1}
\end{table}
\begin{table}[h!]
	\centering
	\renewcommand{\arraystretch}{1.2}
	\begin{tabular}{|c|c|c|c||c|c|c|}
		\hline
		\multicolumn{7}{|c|}{\textbf{Scenario 2}}\\
		\hline
		\multirow{2}{*}{\centering \textbf{Model}}& \multicolumn{3}{c|}{\textbf{MAE}} & \multicolumn{3}{c|}{\textbf{MedAE}} \\ \cmidrule{2-7}
		& $\tau = 0.99$   & $\tau = 0.995$ & $\tau = 0.9995$ & $\tau = 0.99$  & $\tau = 0.995$ & $\tau=0.9995$ \\ \hline
		\textbf{gev}            &    17.9                                                      &             21.0              &          34.8           &   18.1                  &    21.3                     &       36.5                       \\ \hline
		\textbf{grf}            & 18.5                                     &  22.2                          &  83.7                  &  
		
		18.3  & 21.8                                                                & 38.7                        \\ \hline
		\textbf{qrf}            &18.8                                                    &22.9                             &   88.1                &   18.5                          &   22.0                                                                   & 37.4                           \\ \hline
		
	\end{tabular}
	\caption{Performance of models for different metrics.}
	\label{tab:performanceScena2}
\end{table}

\begin{table}[h!]
	\centering
	\renewcommand{\arraystretch}{1.2}
	\begin{tabular}{|c|c|c|c||c|c|c|}
		\hline
		\multicolumn{7}{|c|}{\textbf{Scenario 3}}\\
		\hline
		\multirow{2}{*}{\centering \textbf{Model}}& \multicolumn{3}{c|}{\textbf{MAE}} & \multicolumn{3}{c|}{\textbf{MedAE}} \\ 
		\cmidrule{2-7}
		& $\tau = 0.99$   & $\tau = 0.995$ & $\tau = 0.9995$ & $\tau = 0.99$  & $\tau = 0.995$ & $\tau=0.9995$ \\ \hline
		\textbf{gev}            &                             
		9.50                                                                                 &      11.3                     &      19.7                             & 9.08                                                                                &      10.8                     &      19.3        \\ \hline
		\textbf{grf}            & 10.1                                                                     &  12.7              &           36.2        &  9.85                                                                    &        12.2                 &                            33.7          \\ \hline
		\textbf{qrf}            &  10.3                                                                    & 12.8                &           34.1       &  10.9                     &   11.8                     &                31.7          \\ \hline
		
	\end{tabular}
	\caption{Performance of models for different metrics.}
	\label{tab:performanceScena3}
\end{table}

In the appendix~\ref{addi_simul}, we provide additional graphical representations for the different scenarios considered in this work. These graphs validate our conclusions by illustrating the method's performance for different orders of high quantiles. In so doing, we provide an overview of the efficiency and robustness of our approach in handling extreme quantile regression tasks.

\section{Applications to real datasets}\label{section_real_appli}
To evaluate the performance of the extreme quantile regression model GEV-erf in a real-world setting, we use daily meteorological data from the Fort Collins station in Colorado (USA), recorded between January 1, 1900, and December 31, 1999. This dataset, available in \cite{sielenou_fcwxcsv_2020}, has been used in previous studies such as \cite{dkengne_automatic_2020} and \cite{katz_statistics_2002}. In our analysis, the response variable $Y$ corresponds to the daily maximum temperature (in degrees Fahrenheit), while the covariates include daily accumulated precipitation (in inches), daily accumulated snowfall, and two transformed variables: the variable \textit{Season}, which takes values 1 (winter), 2 (spring), 3 (summer), and 4 (fall), and the maximum temperature of the previous day. Observations with missing values in any of the variables were excluded, resulting in a total of 35,793 complete observations. To increase the dimensionality of the covariate space and test the robustness of the model in a high-dimensional setting, we add six independent random variables generated from a uniform distribution on $[-1,1]$, bringing the total number of covariates to $p=10$. The analysis focuses on monthly maximum temperatures, obtained by forming monthly blocks from the daily data. For model training, $70\%$ of the monthly maxima are used as the training set, with the remaining $30\%$ reserved for testing.

\begin{figure}[h]
	\centering
	\includegraphics[width=0.99\linewidth,height=0.5\textheight]{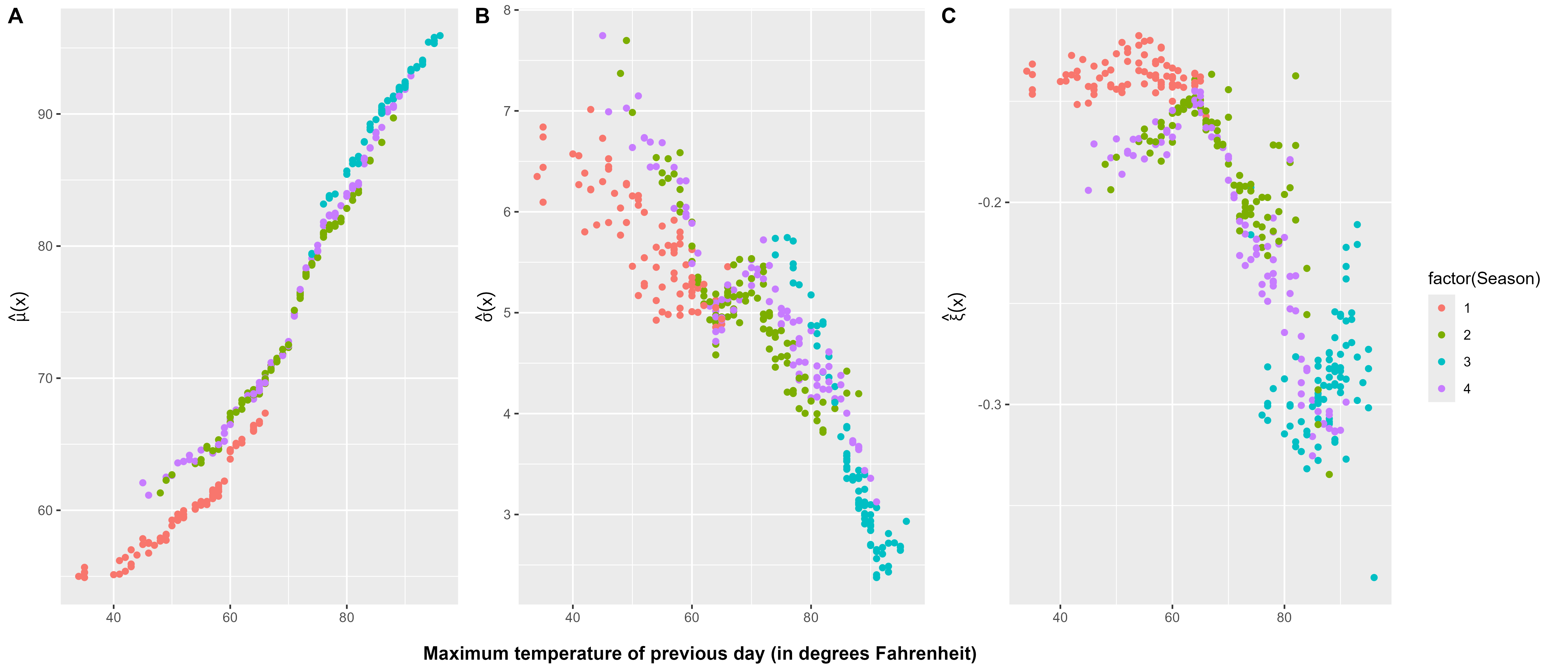}
	\caption{	Variation of the estimated parameters $\hat{\theta}(x)$ as a function of the previous day’s maximum daily temperature, by season.	}
	\label{fig:plotPreviousDay_param}
\end{figure}
\begin{figure}[h]
		\centering
		\includegraphics[width=0.99\linewidth,height=0.5\textheight]{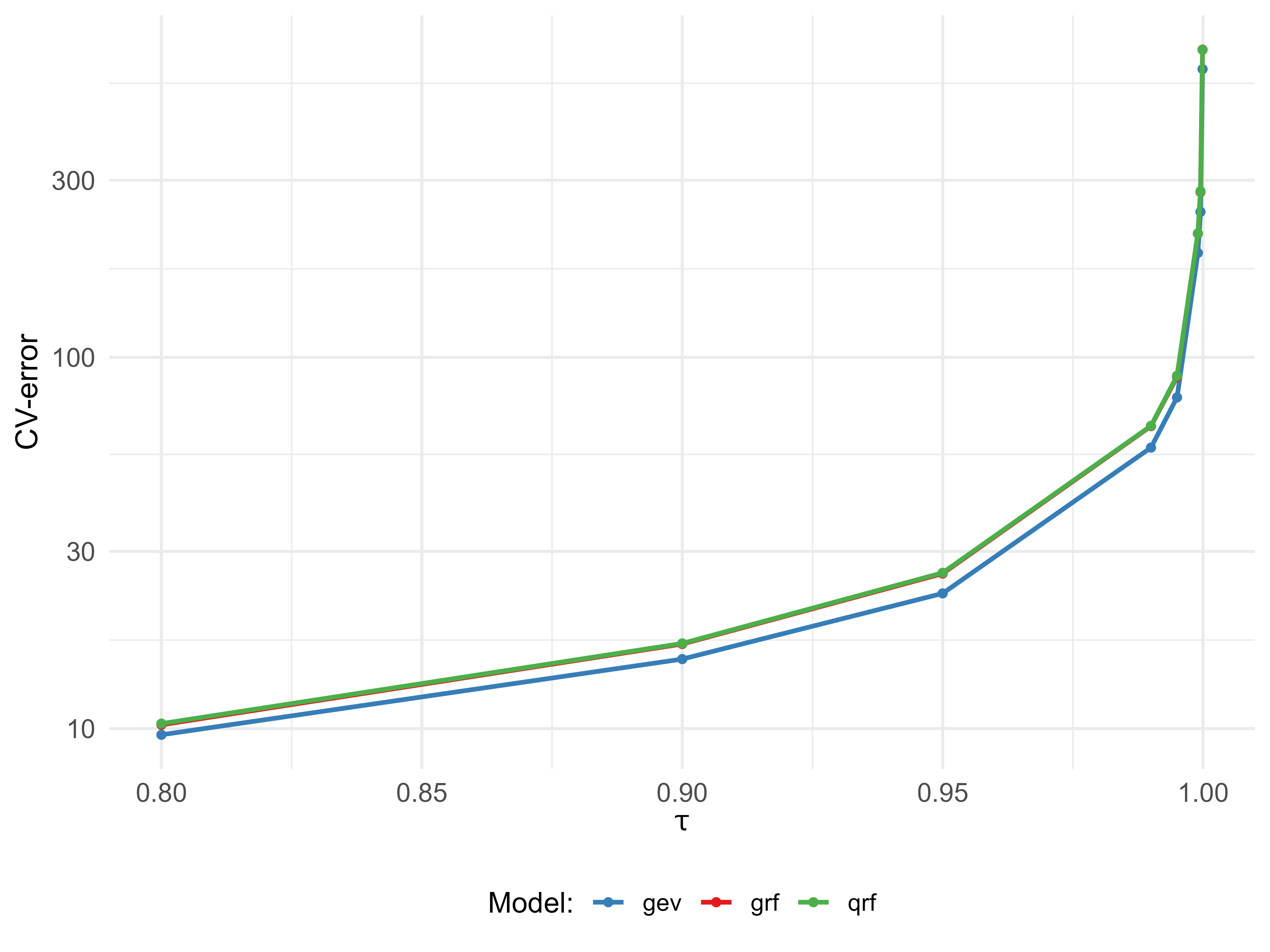}
		\caption{Average prediction error for the different models as a function of the extreme quantile level.}
		\label{fig:ploterrorwangli}
	\end{figure}

The results reveal a strong dependence between the monthly maximum temperature and the previous day’s maximum temperature. Figure~\ref{fig:plotPreviousDay_param} illustrates the variation of the estimated GEV distribution parameters $\hat{\theta}(x)$ with respect to this covariate, with a distinction made across seasons. A nearly linear increasing relationship is observed between the previous day's maximum temperature and the location parameter $\hat{\mu}(x)$, suggesting that warmer preceding days raise the baseline level of expected extreme temperatures. The scale parameter $\hat{\sigma}(x)$ decreases as the previous day's temperature increases, indicating reduced variability in the extremes. The shape parameter $\hat{\xi}(x)$ exhibits a more complex dynamic: for moderate temperatures, $\hat{\xi}(x)$ remains close to zero, indicating a Gumbel-type tail (moderately heavy). However, for higher temperatures—particularly during summer and autumn—$\hat{\xi}(x)$ becomes significantly negative. This implies a shorter tail, suggesting the presence of an upper bound on extreme temperatures. These findings are consistent with the conclusions of \cite{dkengne_automatic_2020}, who showed that the unconditional distribution of daily maximum temperature belongs to the Weibull domain of attraction ($\xi < 0$), implying a finite right endpoint. Our conditional modeling using the \texttt{GEV-erf} model confirms this property, with estimated values of the shape parameter ranging between $-0.386$ and $-0.118$, as shown in panel C of Figure~\ref{fig:plotPreviousDay_param}.	
We conduct a quantitative performance evaluation of the \texttt{GEV-erf} method by comparing it with the \texttt{GRF} approach of \cite{athey_generalized_2019} and the \texttt{QRF} method of \cite{meinshausen_quantile_2006}, using the metric proposed by \cite{wang_estimation_2013}, defined as follows:
	\begin{equation} \label{WangMetrique}
	R_{n'}\left( \hat{Q}_{.}(\tau) \right) = \frac{\sum_{i=1}^{n'} \mathbb{1} \{ Y_i < \hat{Q}_{X_i}(\tau) \} - n'\tau}{\sqrt{n'\tau (1 - \tau)}},
	\end{equation}	
where the function $\hat{Q}_x(\tau)$ denotes the estimated conditional quantile, evaluated on the test sample ${(x_i, y_i)}_{i=1,\ldots,n'}$. The predictive performance of the different models is estimated using 5-fold cross-validation. As an evaluation criterion, we use the absolute value of the metric  defined in (\ref{WangMetrique}). This procedure is repeated 20 times, and the average of the resulting errors is computed to stabilize the evaluation. Figure~\ref{fig:ploterrorwangli} illustrates the evolution of the average prediction error as a function of the probability level (on a logarithmic scale) for each of the evaluated models. The results clearly show that the \texttt{GEV-erf} method outperforms the \texttt{GRF} and \texttt{QRF} approaches, the latter two exhibiting very similar performance regardless of the extreme probability level considered.
\section{Conclusion}
Extreme quantile regression is a powerful statistical tool that allows for the analysis and prediction of behaviours in the tails of distributions, where rare and extreme events occur. The existing literature on modeling methods for this approach is limited, particularly in the context of using the Block Maxima approach to ensure tail extrapolation. Most available methods use the Peak-Over-Threshold approach (\cite{pasche_neural_2023}, \cite{velthoen_gradient_2023}, \cite{farkas_generalized_2024}, \cite{gnecco_extremal_2024}). In this work, we propose a flexible quantile regression method based on the BM approach  and the generalized random forest method to address the issues encountered with classical quantile regression methods, primarily the limitation to low-dimensional covariate spaces and the potential complexity of the quantile function. Our proposed method effectively addresses these issues. Using the BM approach, we model the tail of the conditional distribution $Y|X=x$ for the generalized extreme value distribution with parameters dependent on covariates. These parameters are estimated using a penalized weighted maximum likelihood method, with weights obtained through the generalized random forest method. Simulation studies and  application to real dataset show that our method better captures the complex structure of the quantile function and provides good estimates even when the characteristic variables are high-dimensional and  in the presence of noise.  Our method demonstrates strong performance compared to other quantile regression methods that utilize learning algorithms, such as \cite{meinshausen_quantile_2006} and \cite{athey_generalized_2019} method.
While this work is more application-oriented, future research should aim to theoretically prove the consistence of our method, as done by \cite{gnecco_extremal_2024} for the POT approach. Another perspective is to explore the multidimensional case of Y (i.e., $Y \in \mathcal{Y} \subset \mathbb{R}^q$ with $q > 1$) and examine the spatial aspect of our method.

\backmatter

%
%

\section*{Disclosure Statement}
No potential conflict of interest was reported by the authors.

%


\bibliography{Reference}

\begin{appendices}
		
	\section{Selection of parameters $\lambda$ and min.node.size }\label{choix_hyperpara} 
	The table (\ref{table:parametres}) shows the results of the cross-validation used to select the penalty $\lambda$ and the tuning parameter $min.node.size$ for the generalized random forest. To optimize the execution time of the cross-validation, we set the number of trees in the forest to $num.trees = 2000$ (the default value in the GRF method). The various $\lambda$ values tested are $\{10^{-4}, 10^{-3}, 0.005, 0.01, 0.05, 0.1 \}$ and those of $min.node.size$ are $ \{5 , 10, 20, 50 \} $ using 5-fold cross-validation (i.e. we take $K=5$ in (\ref{CV})). We select the pair $(\lambda, min.node.size)$ that minimizes the cross-validation error defined in (\ref{CV}). The table (\ref{table:parametres}) shows the cross-validation errors for all combinations of these hyper-parameters in the set under consideration, and for the different scenarios performed.  According to this table, the smallest cross-validation error is obtained for the combination ($\lambda=0.001;$ $min.node.size=10$) for scenario 1, ($\lambda=0.001;$ $min.node.size=50$) for scenario 2 and ($\lambda=0.001;$ $min.node.size=5$) for scenario 3.  
	\begin{center}
		\begin{table}[h!]
			\begin{tabular}{|c|c||c|c|c|}
				\hline
				\multicolumn{2}{|c||}{\textbf{Grid parameter for CV}} &  \multicolumn{3}{|c|}{\textbf{CV-Error}} \\\hline
				\textbf{$\lambda$} & \textbf{min.node.size} &\textbf{Scenario 1} & \textbf{Scenario 2} & \textbf{Scenario 3} \\
				\hline \hline
				1e-04 & 5 &11.53729 &13.10445& 10.42870  \\
				\hline
				1e-03 &  5 &11.53557 & 13.10410 &\textbf{10.41969}  \\
				\hline
				5e-03  & 5 &  11.53842& 13.10315& 10.42660  \\
				\hline
				1e-02  & 5 & 11.53971& 13.10379 & 10.42475  \\
				\hline
				5e-02 &  5 & 11.53710& 13.10341 & 10.42564  \\
				\hline
				1e-01 & 5   & 11.54149 & 13.10148  &10.42543  \\
				\hline
				1e-04  & 10  & 11.53785 & 13.09767& 10.42836  \\
				\hline
				1e-03 & 10   & \textbf{11.53538} & 13.09863& 10.42696  \\
				\hline
				5e-03 &  10  &11.53980& 13.09724 & 10.43051 \\
				\hline
				1e-02  & 10  & 11.53842 & 13.09817& 10.42812  \\
				\hline
				5e-02 & 10 & 11.53824& 13.09793 & 10.42688  \\
				\hline
				1e-01 & 10 & 11.53911 & 13.09822 & 10.42852  \\
				\hline
				1e-04  & 20  & 11.53810& 13.09522& 10.43054  \\
				\hline
				1e-03 & 20 & 11.53865& 13.09498& 10.42747  \\
				\hline
				5e-03 &  20 & 11.53897& 13.09506 & 10.42786 \\
				\hline
				1e-02  & 20  & 11.53886& 13.09487& 10.43035  \\
				\hline
				5e-02 & 20 & 11.53847& 13.09440 & 10.42615  \\
				\hline
				1e-01 &  20 & 11.53947& 13.09515 & 10.42804  \\
				\hline
				1e-04 &  50  & 11.54066& 13.09314& 10.42667  \\
				\hline
				1e-03 &  50 & 11.53954 & \textbf{13.09209}  & 10.42739 \\
				\hline
				5e-03 &  50 & 11.53960& 13.09295& 10.42588   \\
				\hline
				1e-02  & 50 & 11.54043 & 13.09330& 10.42886  \\
				\hline
				5e-02  & 50  & 11.54106& 13.09272& 10.42899  \\
				\hline
				1e-01 & 50  & 11.54099 &13.09276&  10.42816  \\
				\hline
			\end{tabular}
			\caption{Adjustment parameters with different combinations of \textit{$\lambda$} and \textit{min.node.size}.}
			\label{table:parametres}
		\end{table}
	\end{center}
	\section{Additional simulation study}\label{addi_simul}
\begin{figure}[h!]
		\centering
		\includegraphics[width=0.991\linewidth,height=0.6\textheight]{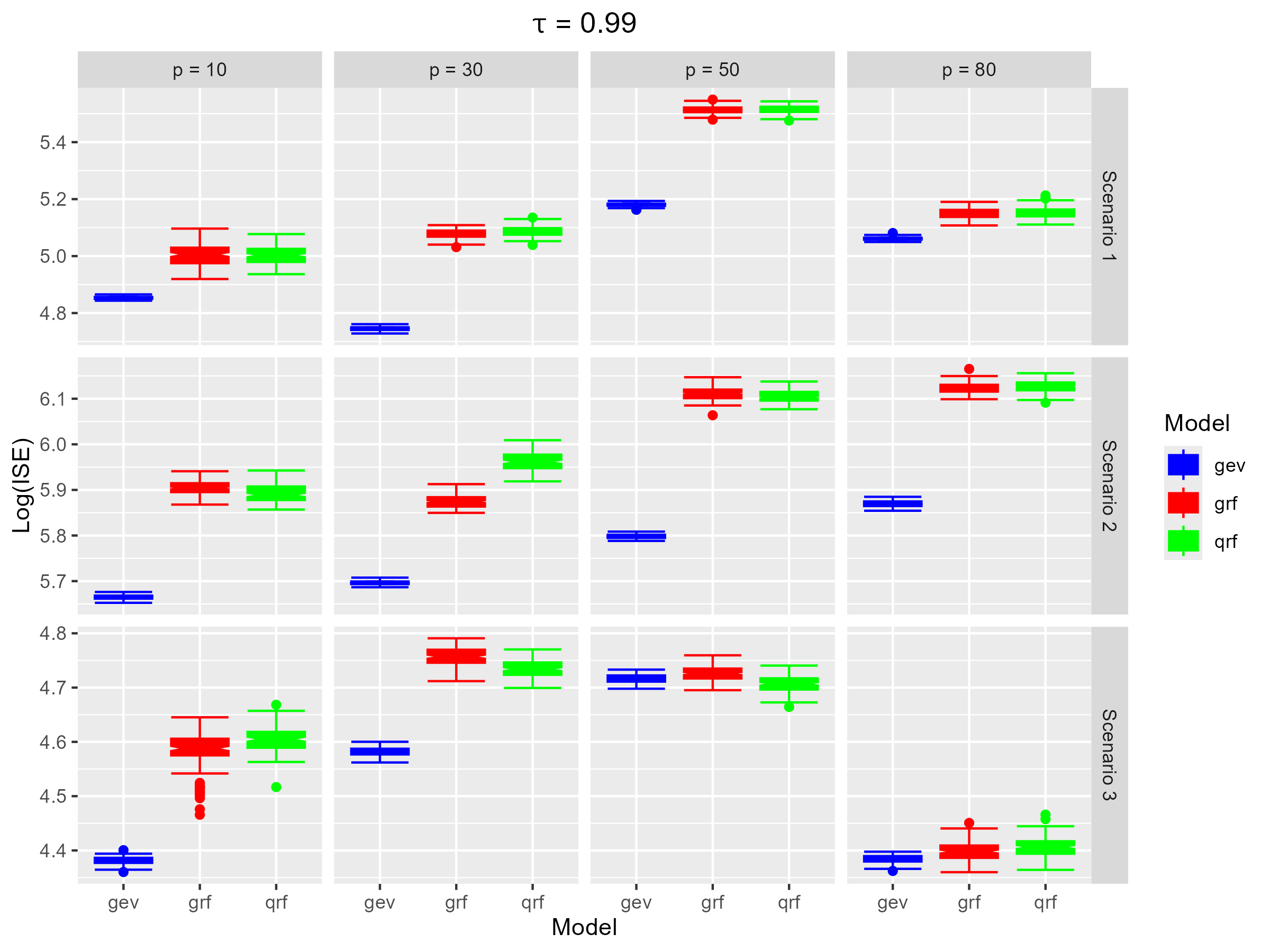}
		\caption{ Boxplot of $log(ISE)$ over $100$ replication, for $p \in \{10,30,50,80\},$ $\tau=0.99$ and different scenario.}
		
		\label{fig:boxplotgridscen13q099}
	\end{figure}

\begin{figure}[h!]
		\centering
		\includegraphics[width=0.99\linewidth,height=0.9\textheight]{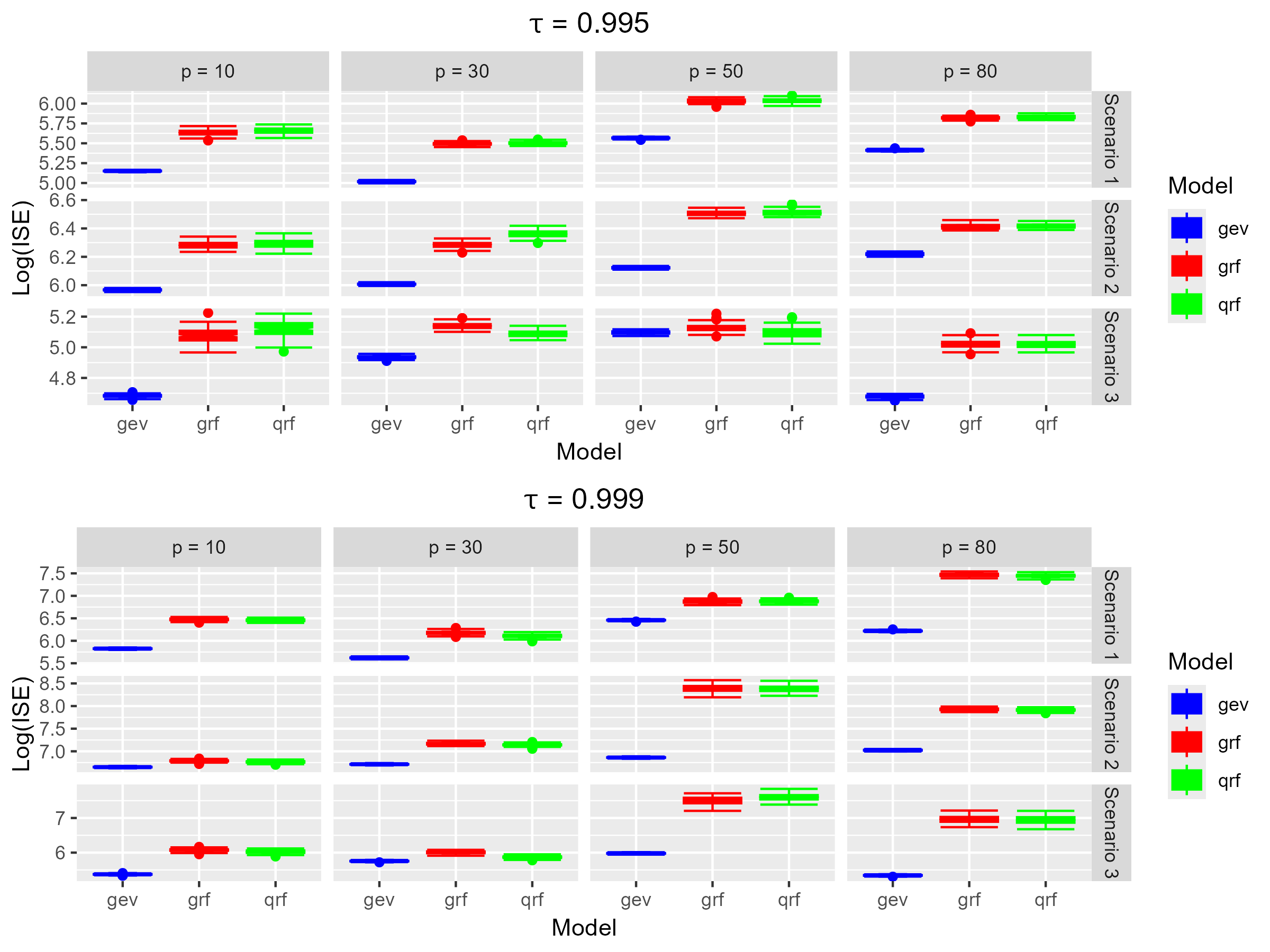}
		\caption{ Boxplot of $log(ISE)$ over $100$ replication, for $p \in \{10,30,50,80\},$ $\tau\in \{0.995,0.999\}$ and different scenario.}
		
		\label{fig:boxplotgridscen13q09950999}
	\end{figure}

\begin{figure}[h!]
		\centering
		\includegraphics[width=0.99\linewidth,height=0.6\textheight]{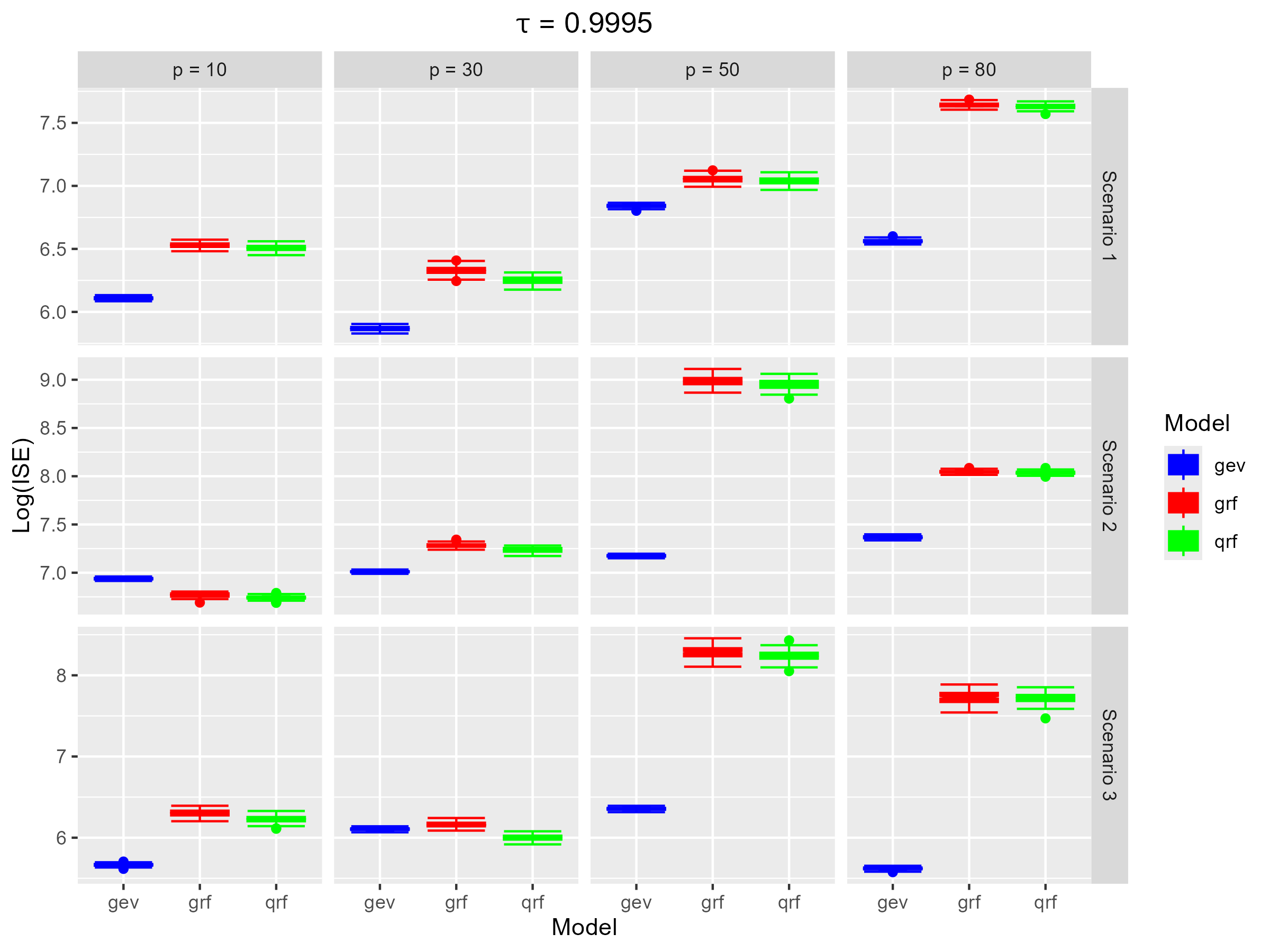}
		\caption{ Boxplot of $log(ISE)$ over $100$ replication, for $p \in \{10,30,50,80\},$ $\tau=0.9995$ and different scenario.}
		\label{fig:boxplotgridscen13q09995}
	\end{figure}

	In this section, we perform additional experiments to demonstrate the performance of our method in different scenarios and for various quantile orders. Whatever the quantile order considered, our model maintains high performance across the various scenarios and covariate sizes studied, as shown in Figures \ref{fig:boxplotgridscen13q099}, \ref{fig:boxplotgridscen13q09950999}, and \ref{fig:boxplotgridscen13q09995}.
	

	\section{ Sensitivity analysis of block size $m$}\label{Sensitivity_analy}
We study here the sensitivity of the block maxima (BM) method to the block size $m$, a crucial parameter for estimating extreme quantiles. An excessively large block size increases the variance of the estimators, while an insufficient block size induces bias. The literature does not provide a universal method for selecting the block size $m$, although several recent contributions have attempted to address this issue. For example, \cite{ozari_new_2019, ozari_proposal_2018} present a computational approach illustrated by a financial case study. \cite{wang_method_2016} propose a multi-criteria method combining graphical analyses and goodness-of-fit tests (Kolmogorov–Smirnov, $\chi^2$). \cite{dkengne_automatic_2020} develop an automatic method applied in engineering and meteorology, while \cite{cervantes_assessing_2024} assess the fit of the GEV distribution across nine representative block sizes using QQ-plots and statistical tests (Kolmogorov–Smirnov, Anderson–Darling, Cramér–von Mises).

Building on these contributions, we analyze the impact of the block size $m$ on both the fit of the estimated GEV distribution and the estimation of the conditional quantile. For each scenario, we consider a range of block sizes from 10 to 100 in increments of 5. For each value of $m$, the model is trained on a dataset of size $N$, and the conditional quantile is then estimated on an independent test sample $\{(x_i, y_i)\}_{i=1}^{n'}$, as described in Section~\ref{sect_Application}. The covariate dimension is fixed at $p = 40$, the regularization parameter at $\lambda = 0.001$, and the parameters \textit{min.node.size} and \textit{num.trees} are retained at their default values from the \texttt{grf} package \cite{athey_generalized_2019}. The GEV-ERF model is fitted to the block maxima samples corresponding to each block size. For each value of $m$, two steps are performed using the test sample. First, the MISE (as defined in Section~\ref{sect_Application}) is computed for various quantile levels ($\tau \in \{0.9, 0.99, 0.999\}$), in order to identify the minimum block size $m_{\min}$ beyond which the MISE stabilizes. Second, for all $m \geq m_{\min}$, we assess the goodness-of-fit of the conditional GEV distribution estimated by the GEV-ERF method using three statistical tests: Kolmogorov–Smirnov (KS), Cramér–von Mises (CVM), and Anderson–Darling (AD). These tests are applied to the probability integral transform of the estimated distribution. More precisely, for each $i \in \{1, \ldots, n'\}$, we compute $u_i = G_{\theta(X_i)}(z_i)$, where $z_i$ is the block maximum from the test sample, and test whether the $u_i$ values follow a uniform distribution.

\begin{figure}[h!]
	\centering
	\includegraphics[width=0.99\linewidth,height=0.5\textheight]{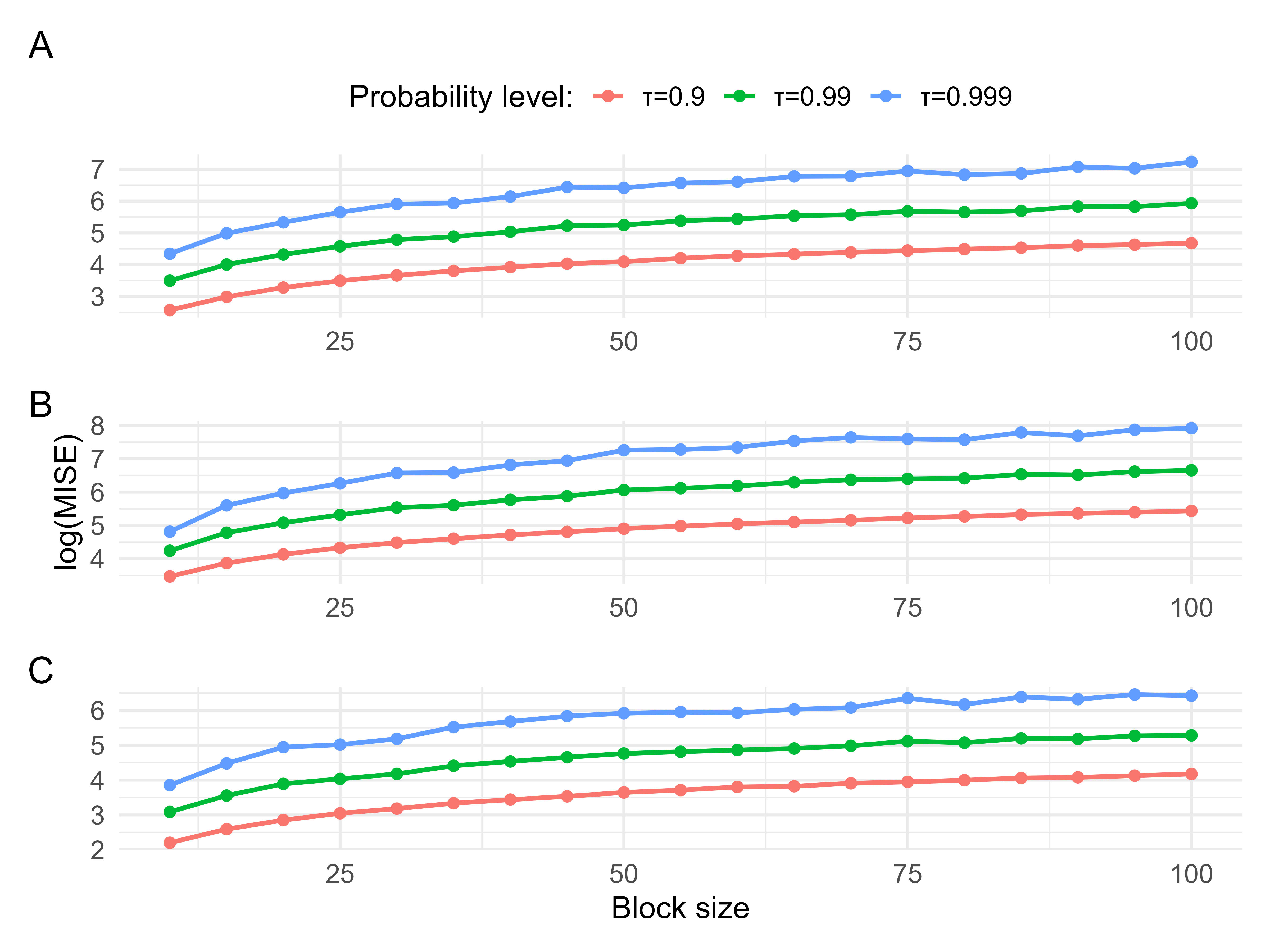}
	\caption{ Log(MISE) of Estimated Conditional Quantiles vs. Block Size for scanrio~1 (A), scenario~2 (B) and scenario~3 (C)}
	\label{fig:log_MisePerBlockSize}
\end{figure}

The goodness-of-fit tests are performed using the functions \texttt{ad.test} (AD test) and \texttt{cvm.test} (CVM test) from the \texttt{goftest} package, and \texttt{ks.test} (KS test) from the \texttt{stats} package. Lower values of the test statistics indicate a better fit between the model and the data. Figure~\ref{fig:log_MisePerBlockSize} shows the evolution of the logarithm of the MISE as a function of block size for various quantile levels across the scenarios. The results indicate that while increasing block size tends to raise the MISE of the estimated quantiles, this increase becomes less variable beyond $m_{\min} = 30$ in Scenarios 1 and 2, and $m_{\min} = 25$ in Scenario 3. This trend can be explained by the fact that smaller blocks provide more observations for training, thereby improving the performance of the estimation method.
 Table~\ref{table_stat_test} presents the KS, AD, and CVM test statistics for block sizes $m \geq m_{\min}$. The smallest values for the KS and CVM statistics occur around $m = 45$ in Scenarios 1 (0.061 and 0.172, respectively) and 2. In Scenario 3, the minimum values across all three tests are achieved at $m = 25$, indicating a good model fit at that block size. These findings confirm that block size significantly influences the quality of fit of the GEV-ERF model. The results highlight the method’s sensitivity to this parameter and suggest that block sizes in the range of 25 to 50 generally provide a good trade-off between bias and variance, ensuring a satisfactory fit between the estimated GEV distribution and the block maxima. The choice of block size remains a central challenge in extreme value analysis. In practice, it is common to select block sizes based on natural time units—such as a year, season, or month—depending on the temporal resolution of the data.

\begin{table}[htbp]
	\centering

	\begin{tabular}{|c|ccc|ccc|ccc|}
	\toprule
		\textbf{Block size} & \multicolumn{3}{|c|}{\textbf{Scenario 1}} & \multicolumn{3}{|c|}{\textbf{Scenario 2}} & \multicolumn{3}{|c|}{\textbf{Scenario 3}} \\
		\cmidrule(lr){2-4} \cmidrule(lr){5-7} \cmidrule(lr){8-10}
		& KS   & AD  & CVM  & KS & AD& CVM  & KS & AD & CVM  \\
		& Stat  &stat & stat & stat &  stat & stat &stat & stat &  \\
		\midrule
25 &  &  &  & & & & 0,053 &1,572 & 0,234\\
30 & 0,072& 1,468 &0,206  & 0,046 & 0,453 & 0,074& 0,080 &1,950 & 0,350\\
35  &0,071&1,290 &0,203 & 0,071& 0,950 & 0,180 & 0,101  &3,845 & 0,728\\
40  &0,077 &1,975 &  0,296&0,055&0,561& 0,088 &0,111 & 3,717 & 0,704  \\
45  &0,061 &1,308&0,172 & 0,036 & 0,237& 0,028 &  0,119 &  4,659 & 0,914 \\
50  &0,073 &1,571 & 0,232& 0,059 & 0,468 & 0,083 &  0,109 & 3,523& 0,648 \\
55 & 0,085&1,753 &0,245 & 0,069 &0,652& 0,103 &0,120 &3,243&  0,573 \\
60 &0,090& 1,492 &0,242&0,055 & 0,324 & 0,035 & 0,103  & 2,672 & 0,491 \\
65 &0,095 &1,686&0,300  & 0,064 & 0,451 & 0,062& 0,139  & 4,437 & 0,848\\
70 &0,090 &1,372&0,223& 0,056 & 0,272 &0,031&0,103 &1,851 &  0,307 \\
75 &0,102 & 1,478& 0,256&0,047 & 0,285 & 0,030 & 0,097 & 2,171 &0,368  \\
80  &0,140 & 2,296& 0,431 &0,091 & 0,980 & 0,159&  0,112 & 2,876 &0,463 \\
85  &0,121 &2,074&0,385&0,142 & 2,956 & 0,442&  0,138   & 3,946 & 0,735  \\   	
90  &0,102 &1,210&0,206&0,058 & 0,309& 0,025 & 0,128  &2,525 & 0,458  \\
95 &0,112 &1,647&0,259  & 0,058 & 0,281 & 0,023 &  0,149 & 4,444&0,849  \\
100  &0,132 &2,118 &0,388& 0,113 &1,380 & 0,176& 0,124& 1,773 & 0,306  \\
		\bottomrule
	\end{tabular}
	\caption{Statistical values for the various tests as a function of block size (verifying $m\geq m_{min}$ ) and scenario.}
	
	\label{table_stat_test}
\end{table}

\end{appendices}

\end{document}